\documentclass[preprint]{aastex}

\usepackage{graphicx}

\newcommand{\Mo}{M_{\odot}}

\newcommand{\kms}{km~s$^{-1}$}

\newcommand{\bey}{\begin{eqnarray}} \newcommand{\eey}{\end{eqnarray}} 
\newcommand{\beq}{\begin{equation}} \newcommand{\eeq}{\end{equation}}

\received{ }
\revised{ }
\accepted{ }

\begin{document}

\title{Dynamical Evolution of the Mass Function of Globular Star Clusters}

\author{
S. MICHAEL FALL\altaffilmark{1}  AND
QING ZHANG\altaffilmark{1,2} 
}

\email{fall@stsci.edu, qzhang@stsci.edu}

\altaffiltext{1}{Space Telescope Science Institute, 3700 San Martin Drive,
                 Baltimore, MD 21218}
\altaffiltext{2}{Department of Physics and Astronomy, Johns Hopkins
                 University, 3400 N. Charles Street, Baltimore, MD 21218}

\begin{abstract}

We present a series of simple, largely analytical models to compute 
the effects of disruption on the mass function of star clusters. 
Our calculations include evaporation by two-body relaxation and 
gravitational shocks and mass loss by stellar evolution. 
We find that, for a wide variety of initial conditions, the mass 
function develops a turnover or peak and that, after 12 Gyr, this 
is remarkably close to the observed peak for globular clusters, at 
$M_p\approx2\times10^5~\Mo$. 
Below the peak, the evolution is dominated by two-body relaxation, 
and the mass function always develops a tail of the form $\psi(M) = 
{\rm const}$, reflecting that the masses of tidally limited clusters 
decrease linearly with time just before they are destroyed. 
This also agrees well with the empirical mass function of globular 
clusters in the Milky Way.
Above the peak, the evolution is dominated by stellar evolution at 
early times and by gravitational shocks at late times. 
These processes shift the mass function to lower masses while nearly
preserving its shape.
The radial variation of the mass function within a galaxy depends 
on the initial position-velocity distribution of the clusters.
We find that some radial anisotropy in the initial velocity distribution,
especially when this increases outward, is needed to account for the 
observed near-uniformity of the mass functions of globular clusters. 
This may be consistent with the observed near-isotropy of the present 
velocity distributions because clusters on elongated orbits are 
preferentially destroyed.
These results are based on models with static, spherical galactic
potentials.  
We point out that there would be even more radial mixing of the orbits
and hence more uniformity of the mass function if the galactic potentials
were time-dependent and/or non-spherical.

\end{abstract}


\section{Introduction} 

Globular clusters appear to have a preferred mass scale.
Their mass function has a turnover or peak at $M_p \approx 2\times 
10^5~\Mo$ and a dispersion of only $\sigma(\log M) \approx 0.5$ 
(Harris 1991).
The mass and luminosity functions of globular clusters are often
modeled as lognormal functions, although they can also be represented 
by several broken power laws (McLaughlin 1994).
Moreover, globular clusters represent the most numerous gravitationally 
bound stellar subsystems within the spheroidal components of galaxies 
(the others being dwarf galaxies).
There are relatively few subsystems in galactic spheroids with 
masses between those of individual stars and globular clusters 
(i.e., in the range $10^0 \lesssim M \lesssim 10^4~\Mo$) and between 
those of globular clusters and the spheroids themselves (i.e., in the 
range $10^6 \lesssim M \lesssim 10^{10}~\Mo$ for a large galaxy). 
The preferred mass scale of old globular clusters may be an important 
clue in understanding their formation and evolution. 
The corresponding feature in the luminosity function, at $M_V
\approx -7.3$, is sometimes used as a distance indicator (e.g.,
Whitmore 1997).

In contrast, the mass functions of many other types of astronomical 
objects appear to be scale-free and are often modeled as single 
power laws. 
For diffuse and molecular clouds in the Milky Way, the mass function has 
the form $\psi(M) \propto M^{\beta}$ with $\beta \approx -2$ over the 
range $10^{-1}\lesssim M \lesssim 10^6~\Mo$ (Dickey \& Garwood 1989; 
Solomon \& Rivolo 1989).
For young star clusters in the disks of normal galaxies (i.e., open 
clusters), the luminosity functions, which may or may not reflect the 
mass functions, are power laws, $\phi(L) \propto L^{\alpha}$ with 
$\alpha\approx -2$ (Milky Way: van den Bergh \& Lafontaine 1984; Large 
Magellanic Cloud: Elson \& Fall 1985; M33: Christian \& Schommer 1988).
Finally, for the young star clusters formed in interacting and merging
galaxies, the luminosity and mass functions also have power-law form,
with $\alpha \approx \beta \approx -2$ for $10^4\lesssim M\lesssim 
10^6~\Mo$ (Whitmore et al. 1999; Zhang \& Fall 1999, and references 
therein
\footnote{Whitmore et al. (1999) found that a double power law with 
$\alpha_1=-1.7$ and $\alpha_2 = -2.6$ provided a slightly better fit
than a single power law with $\alpha\approx -2$ to the luminosity 
function of young star clusters in the Antennae galaxies 
(NGC 4038/9). 
However, this still contrasts sharply with the luminosity function 
of old globular clusters. 
Suggestions that the luminosity and mass functions of the young star 
clusters in the Antennae might be similar to those of old globular 
clusters were based on earlier, less sensitive observations, which could 
not reach beyond the putative turnovers in these functions (Meurer 1995; 
Fritze-von Alvensleben 1999).}).
The last finding is particularly significant because these clusters
are often regarded as young globular clusters.

Two explanations have been proposed for the preferred mass scale of
old globular clusters. 
One is that the conditions in ancient galaxies and protogalaxies 
favored the formation of objects with masses $\sim 10^5$--$10^6~\Mo$ 
but that these conditions no longer prevail in modern galaxies. 
For example, the minimum mass of newly formed star clusters, set by 
the Jeans mass of interstellar clouds, will be high when the gas cannot 
cool efficiently and low when it can, which in turn will depend on the 
abundances of heavy elements and molecules, the strength of any heat 
sources, and so forth. 
These effects may have suppressed the formation of low-mass clusters 
in the past but not at present (Fall \& Rees 1985; Kang et al. 1990).
The other explanation for the preferred mass scale of old globular 
clusters is that they were born with a much wider spectrum of masses 
that was later modified by the selective destruction of low-mass clusters 
(Fall \& Rees 1977; Gnedin \& Ostriker 1997, and references therein).
In this case, a power-law mass function might evolve into a 
lognormal-like mass function (Vesperini 1997, 1998; Baumgardt 1998).
This idea is appealing because the masses and sizes of the brightest
young clusters in merging galaxies are similar to those of the old 
globular clusters in the spheroids of galaxies.

Star clusters are relatively weakly bound objects and are vulnerable 
to disruption by a variety of processes that operate on different 
timescales.
Stellar evolutionary processes remove mass from clusters by a
combination of supernovae, stellar winds, and other ejecta.
These are effective on both short timescales ($t\lesssim 10^7$ yr), 
when the clusters or protoclusters are partly gaseous, and on
intermediate timescales ($10^7\lesssim t\lesssim {\rm few} 
\times 10^8$ yr), when the clusters consist entirely of stars.
Three stellar dynamical processes remove mass from clusters on 
long timescales ($t\gtrsim {\rm few} \times 10^8$ yr). 
First, internal relaxation by two-body scattering causes some stars 
to gain enough energy to escape from the clusters. 
Second, as clusters orbit around a galaxy, they experience a 
time-dependent tidal field, which may vary rapidly enough when they
pass near the galactic bulge or through the galactic disk that stars
in the outer parts of the clusters cannot respond adiabatically.
The corresponding changes in the energy of the stars (heating
and relaxation) cause some of them to escape.
These effects are known respectively as bulge and disk shocks and
more generically as gravitational shocks.
Third, dynamical friction, the deceleration of clusters induced by 
the wakes of field stars or dark matter particles behind them, causes 
the clusters to spiral toward the galactic center, where they may be 
destroyed by the strong tidal field.

A potentially serious problem with the idea that disruption
causes the turnover in the mass function of globular clusters is that
the chief disruptive processes operate at different rates in different
parts and different types of galaxies (Caputo \& Castellani 1984;
Chernoff, Kochanek, \& Shapiro 1986; Chernoff \& Shapiro 1987;
Aguilar, Hut, \& Ostriker 1988; Gnedin \& Ostriker 1997; Murali
\& Weinberg 1997a,b,c). 
For example, the rate at which stars escape by two-body relaxation 
depends on the density of a cluster, which is determined by the tidal 
field, and hence is higher in the inner parts of galaxies than in the 
outer parts.
The rate at which stars escape by gravitational shocks is also higher 
in the inner parts of galaxies, both because the orbital periods are 
shorter there and because the surface density of the disk is higher 
there. 
Moreover, disks are absent in elliptical galaxies.
Thus, if the mass function were strongly affected by disruptive 
processes, one might expect its form to depend on radius within a 
galaxy and to vary from one galaxy to another. 
This, however, appears to be contradicted by many observations showing
that the luminosity function of clusters (a mirror of the mass function
when the spread in ages is relatively small) varies little, if at all,
within and among galaxies (Harris 1991). 

The goal of this paper is to explore the evolution of the mass 
function of star clusters by a variety of disruptive processes, 
including evaporation by two-body relaxation and gravitational 
shocks and mass loss by stellar evolution. 
We are especially interested in how the mass function is affected 
by different position-velocity distributions of the clusters, and 
which of these are compatible with observations.
We formulate this problem in terms of a simple, approximate model
that can be solved largely analytically. 
This clarifies how the mass function is affected by the different 
disruptive processes and different position-velocity distributions.
Our calculations are performed in the context of static, spherical
galactic potentials. 
But we also discuss qualitatively how our results would be affected 
by time-dependent and non-spherical galactic potentials.
The plan for the remainder of the paper is as follows.
In Section~2, we specify our model, with the associated assumptions, 
equations, and parameters. 
We present the results of our calculations in Section~3, showing the 
influence of each physical effect on the evolution of the mass function.
In Section~4, we compare our results with previous studies, and in 
Section~5, we summarize our conclusions.

\section{Models} 

We are interested here in the evolution of the mass function of 
star clusters, defined such that $\psi(M,t)dM$ is the number of 
clusters with masses between $M$ and $M+dM$ at time $t$. 
We assume that the clusters present initially, at $t=0$, lose 
mass continuously (smoothing over the escape of individual stars) 
and that no clusters are created subsequently.
Then the mass function must satisfy the continuity equation 
\beq 
\frac{\partial\psi}{\partial t}+ \frac{\partial} {\partial M}
    (\psi \dot{M}) = 0,
\eeq
where the dot denotes differentiation with respect to $t$. 
The formal solution of this equation is 
\beq
\psi(M,t) = \psi_0(M_0) |\partial M_0/\partial M|,
\eeq
where $\psi_0(M) = \psi(M,0)$ is the initial mass function, and 
$M_0(M,t)$ is the initial mass of a cluster that has a mass $M$
at a later time $t$. 
In addition to these variables, the mass function may depend on the
orbital parameters of the clusters, and hence their location within 
a galaxy, and may also vary from one galaxy to another.
We first consider clusters on the same orbit.
Later, we will average the mass function over realistic distributions
of orbits and examine its dependence on the properties of the host 
galaxy.

We consider three processes that reduce the masses of star clusters:
evaporation driven by two-body relaxation, evaporation driven by 
gravitational shocks, and mass loss driven by stellar evolution 
(supernovae, stellar winds, and other ejecta). 
Dynamical friction, combined with tidal limitation, also causes 
disruption, but this is only important near the centers of galaxies
and is neglected here, mainly to simplify our analysis.
More specifically, a cluster of mass $M$ at a distance $R$ from the 
center of a galaxy with a circular velocity $V_c$ would be destroyed 
in a time $t_{df} \approx 60(V_c/220~{\rm km~s}^{-1})(M/10^5~\Mo)^{-1} 
(R/{\rm kpc})^2$~Gyr (see equation 7-26 of Binney \& Tremaine 1987).
We have evaluated this expression for the 146 globular clusters in the 
Milky Way with known luminosities (assuming $M/L_V=3$) and positions 
in the most recent compilation of data by Harris (1996, 1999).
We find that only two clusters have $t_{df} < 10$~Gyr and only seven 
have $t_{df} < 20$~Gyr; the vast majority have much larger $t_{df}$
and are thus virtually immune to disruption by dynamical friction. 
This suggests, but does not prove, that dynamical friction was also 
relatively unimportant in the past.
Bulge shocks contribute to the disruption of clusters on highly 
elongated orbits, but they are probably less important than disk 
shocks for most clusters and are also neglected here.
The rates of evaporation by bulge and disk shocks depend on the 
properties of the clusters and their orbits in similar ways.
Thus, by including strong disk shocks near the centers of galaxies,
we also mimic at least qualitatively the effects of bulge shocks.

As an approximation, we assume that the processes considered 
here---two-body relaxation, gravitational shocks, and stellar 
evolution---operate independently of each other, at fractional 
rates $\nu_{ev}$, $\nu_{sh}$, and $\nu_{se}$. 
Thus, following many previous studies (see Spitzer 1987), we write 
\beq
\dot{M} = -(\nu_{ev}+\nu_{sh}+\nu_{se})M,
\eeq
with
\beq
\nu_{ev}=\frac{\xi_e}{t_{rh}} 
        =\frac{7.25\xi_e mG^{1/2}\ln\Lambda}{M^{1/2}r_{h}^{3/2}}, 
\eeq
\beq
\nu_{sh}=\frac{7\kappa_s\bar{A}}{3t_{sh}}
        =\frac{15.6\kappa_s\bar{A}g_{m}^2r_h^3}{GMP_{\phi}V_Z^{2}}.
\eeq
In equation~(4), $\xi_e$ is the fraction of stars that escape 
per half-mass relaxation time $t_{rh}$ by two-body scattering,
$r_h$ is the half-mass radius, $m$ is the mean stellar mass, and  
$\ln\Lambda$ is the Coulomb logarithm.
We neglect possible slow variations in the last two quantities and 
set $m = 0.7~\Mo$ and $\ln\Lambda = 12$ in all our calculations.
In equation~(5), $t_{sh}$ is the gravitational shock heating time 
for first-order energy changes in the impulse approximation 
(Ostriker, Spitzer, \& Chevalier 1972) and $\bar{A}$ is a correction 
for partial adiabatic (i.e., non-impulsive) response averaged over 
all the stars in a cluster (see Appendix~A for details).
The factor $7/3$ accounts approximately for the addition of 
second-order energy changes, also known as shock-induced
relaxation (Spitzer \& Chevalier 1973; Kundi\'c \& Ostriker 1995).
The other coefficient in equation~(5) relates the fractional change 
in energy caused by gravitational shocks to the corresponding
fractional change in mass, i.e., $\dot{M}/M = \kappa_s \dot{E}/E$.
Also in equation~(5), $V_Z$ is the vertical component of the velocity 
of a cluster relative to the galactic disk, $P_{\phi}$ is the azimuthal 
period of its orbit around the galaxy, and $g_m = 2\pi G\Sigma_d$ 
is the maximum vertical acceleration caused by the disk of surface
mass density $\Sigma_d$.
We assume that the disk has an exponential profile all the way into 
the galactic center, $\Sigma_d (R)= \Sigma_d (0) \exp(-R/R_d)$, thus
helping to mimic the effects of bulge shocks. 
The fractional rate of mass loss by stellar evolution depends on 
the age of a cluster and the stellar initial mass function (IMF). 
We compute $\nu_{se}(t)$ from the Leitherer et al. (1999) models 
with the Salpeter IMF.

We assume each cluster has an outer, limiting radius $r_t$ 
determined by the tidal field of the host galaxy at the pericenter 
of its orbit.
Clusters on orbits with fixed pericenters, such as those in a
static, spherical galactic potential, as assumed here, will 
therefore evolve at constant mean density, $\bar{\rho} = 
M/(4\pi r_t^3/3)$; and clusters on orbits with different 
pericenters will have different $\bar{\rho}$ (with an additional 
weak dependence on the shape of the orbits, resulting from the 
centrifugal acceleration at pericenter; see equation (15) below).
This is a standard assumption, although it is not expected to be
perfect except possibly for circular orbits (Spitzer 1987).
The assumption that the clusters are tidally limited is justified 
by the fact that, if they initially extended beyond $r_t$, their 
outer parts would be stripped off after a few orbits, whereas if 
they did not initially extend to $r_t$, they would expand as a 
result of the disruptive effects considered here, predominantly 
stellar mass loss in the early stages, until they reached $r_t$.
Some clusters with low central concentrations might be destroyed
relatively quickly by a combination of stellar evolution and tidal 
limitation, with little or no help from two-body relaxation or 
gravitational shocks (Chernoff \& Weinberg 1990; Fukushige \& 
Heggie 1995); these clusters are not included in our calculations.

We must now specify the escape probability parameter for 
two-body relaxation $\xi_e$, the energy-mass conversion factor 
for gravitational shocks $\kappa_s$, and the relation between 
the half-mass and tidal radii, $r_h$ and $r_t$.
A valuable point of reference is H\'enon's (1961) model for the 
self-similar evolution of a tidally limited cluster with a single 
stellar mass by two-body relaxation alone. 
This has $\xi_e = 0.045$ and $r_h=0.145 r_t$.
The H\'enon model is often regarded as an adequate approximation 
for high-concentration clusters before core collapse, which 
typically occurs about half way through their lifetimes, and an 
excellent approximation for all clusters after core collapse.
The value of $\xi_e$ found in Monte Carlo and Fokker Planck models 
with a single stellar mass is typically 2--3 times below the H\'enon 
value in the early stages of evolution and closer to it in the late 
stages (Spitzer \& Chevalier 1973; Lee \& Ostriker 1987; Gnedin,
Lee, \& Ostriker 1999).
On the other hand, models with a realistic spectrum of stellar 
masses evolve a few times faster than those with a single stellar
mass (Johnstone 1993; Lee \& Goodman 1995).
Thus, in most of our calculations, we adopt the H\'enon value of 
$\xi_e$ as a reasonable approximation to the effective escape 
probability parameter for the entire evolution of a realistic
cluster, including both its pre- and post-core collapse phases.
We also adopt the relation between $r_h$ and $r_t$ in the H\'enon 
model.

The energy-mass conversion factor $\kappa_s$ depends on how the
energy imparted to a cluster by gravitational shocks is divided 
between bound and escaping stars.
The detailed, but non-evolutionary calculations by Chernoff et al.
(1986) give $\kappa_s \approx 1$ for high-concentration clusters 
($\kappa_s$ is $-\nu/f$ in their notation).
In the evolving Monte Carlo models of Spitzer \& Chevalier (1973), 
which include two-body relaxation and the first- and second-order 
energy changes caused by impulsive gravitational shocks, the total 
evaporation rate is given to an accuracy of about 30\% by equation~(3) 
with $\nu_{sh} = 2/t_{sh}$, corresponding to $\kappa_s \approx 1$ 
(see also Section 5-2b of Spitzer 1987).
Recent Fokker-Planck calculations indicate that the rates of 
evaporation by two-body relaxation and gravitational shocks, 
$\nu_{ev}$ and $\nu_{sh}$, are sometimes correlated and mutually 
reinforcing (Gnedin et al. 1999).
The effect appears to be important mainly when the two rates are 
comparable. 
We neglect this complication and adopt $\kappa_s = 1$.
This and our adopted value of $\xi_e$ ensure that our simple model 
has the correct behavior when either two-body relaxation or 
gravitational shocks dominate, i.e., in the limits $\nu_{ev} \gg 
\nu_{sh}$ and $\nu_{ev} \ll \nu_{sh}$, and thus when the masses of 
clusters are either small or large (relative to $M_p$).
As we show later, these limiting cases play a major role in 
determining the shape of the mass function of the clusters. 

With the assumptions discussed above, equations (3)--(5) take the form 
\beq
\dot{M} = -\mu_{ev} - (\nu_{sh}+\nu_{se})M,
\eeq
where 
\beq
\mu_{ev}=269\ \xi_e (G\bar{\rho})^{1/2}m\ln\Lambda,
\eeq
\beq
\nu_{sh}=\frac{0.0113\kappa_s\bar{A}g_m^2}{G\bar{\rho}P_{\phi}V_Z^2}
\eeq 
are constants and $\nu_{se}$ is a function of time. 
This has the exact solution 
\beq
M = \{M_0-\mu_{ev}\int_0^t \exp [\nu_{sh}t^{\prime}+S(t^{\prime})]
    dt^{\prime}\} \exp [-\nu_{sh}t-S(t)],
\eeq
with 
\beq S(t)=\int_0^t \nu_{se}(t^{\prime})dt^{\prime}.
\eeq
Equation~(9) can can be inverted to obtain $M_0(M,t)$, and this 
can then be substituted into equation (2) to obtain $\psi(M,t)$ 
for any specified initial mass function $\psi_0(M)$.

Figure~1 shows the evolution of the mass $M$ predicted by 
equations~(7)--(10) for three clusters with different initial 
masses $M_0$ on the same orbit (with the parameters specified in 
the caption).
In the early stages ($t \lesssim 3\times10^8$~yr), the mass drops 
approximately exponentially with time as a result of stellar 
evolution until it reaches about 60\% of its initial value.
Thereafter, the mass declines exponentially with time as a 
result of gravitational shocks and linearly with time as a 
result of two-body relaxation.
Gravitational shocks become relatively less important with 
decreasing mass, and two-body relaxation always dominates in 
the late stages as the mass approaches zero.
The evolution predicted by this simple analytical model is generally
similar to that found in the more accurate Monte Carlo, Fokker-Planck, 
and N-body models, although there are differences in detail (see the
references cited above in connection with the parameters $\xi_e$ and
$\kappa_s$).
However, even the most sophisticated models still involve some
important idealizations, and they sometimes differ from each other 
by as much as they differ from our simple model.
These comparisons indicate that the approximate evolution specified 
by equations~(7)--(10) is suitable for our purposes. 

It is worth pausing here to consider separately the influence of 
each disruptive effect on the mass function. 
This is simplest for a set of clusters on the same galactic orbit
(i.e., with the same values of $\mu_{ev}$ and $\nu_{sh}$).
Inserting equations (9) and (10) with $\nu_{sh}=\nu_{se}=0$, 
$\mu_{ev}=\nu_{se}=0$, and $\mu_{ev}=\nu_{sh}=0$ into equation~(2)
gives 
\bey
\psi(M,t)=\psi_0 (M + \mu_{ev} t) &{\rm two-body\ relaxation\ alone,}\\
\psi(M,t)=e^{\nu_{sh}t}\psi_0(M e^{\nu_{sh}t})&{\rm gravitational\ shocks\ 
          alone,}\\
\psi(M,t)=e^{S(t)}\psi_0 (M e^{S(t)}) &{\rm stellar\ evolution\ alone}.
\eey
For two-body relaxation alone, the masses of clusters decrease 
linearly with time.
This flattens the mass function at low masses but has little 
effect on its shape at high masses, i.e., $\psi(M,t)={\rm const}$ 
for $M\lesssim \mu_{ev}t$ and $\psi(M,t)=\psi_0 (M)$ for 
$M\gtrsim \mu_{ev}t$.
Thus, if the mass function is initially a power law, $\psi_0(M)
\propto M^{\beta}$ with $\beta < 0$, it will develop a bend at 
$M\approx \mu_{ev}t$.
In contrast, for gravitational shocks or stellar evolution alone, 
the masses of clusters decrease exponentially or approximately
exponentially with time.
This preserves the shape of the mass function in the sense that 
both $\psi$ and $M$ are simply rescaled by time-dependent factors. 
Thus, if the mass function is initially a power law, it will 
remain one at all later times: $\psi(M,t)\propto \psi_0(M) 
\propto M^{\beta}$.
In the case of gravitational shocks, the rescaling factor increases 
indefinitely,  whereas in the case of stellar evolution, it saturates 
at $\exp S(t) \approx 1.6$ for $t \gtrsim 3\times 10^8$~yr. 
As we show later, the scaling relations derived here are 
approximately correct even when the mass function is averaged
over realistic distributions of orbits (i.e., with different
values of $\mu_{ev}$ and $\nu_{sh}$).

We now consider clusters on different orbits within a galaxy. 
This is assumed for simplicity to have a static, spherically 
symmetric potential $\Phi(R)$, where $R$ denotes the galactocentric 
radius in spherical (not cylindrical) polar coordinates.
Thus, we can characterize each orbit by the constant values of 
the energy $E$ and angular momentum $J$ per unit mass and the angle 
$\theta$ between the normals of the orbital plane and the disk.
For some purposes, it is useful to reexpress $E$ and $J$ in terms 
of the pericenter and apocenter of the orbit, $R_p$ and $R_a$, and 
the radius $R_c$ of a circular orbit with the same energy
\beq
E=\onehalf V_c^2(R_c) + \Phi(R_c) = \onehalf (J/R_{p,a})^2 + \Phi(R_{p,a}),
\eeq  
where $V_c$ is the circular velocity. 
We further assume that the mass distribution of the galaxy can be 
modeled as a singular isothermal sphere, with $\Phi(R)=V_c^2\ln R$ 
and $V_c = {\rm const}$. 
In this case, the mean density of a cluster is given by
\beq
\bar{\rho}(E,J) = \frac{1}{\pi G} \Big(\frac{3}{2}\Big)^4 
    \Big[1-\ln\Big(\frac{R_p}{R_c}\Big)\Big] \Big(\frac{V_c}{R_p}\Big)^2.
\eeq
We note that this depends mainly on the pericenter of the orbit and 
only weakly (logarithmically) on the shape of the orbit.
Equation~(15) is based on the formula for the tidal radius advocated 
by Innanen, Harris, \& Webbink (1983), which accounts approximately 
for the tidal elongation of clusters. 
The mean density would be reduced by the factor $(2/3)^3$ if the King 
(1962) formula for $r_t$ were adopted.
The precise form of $\bar{\rho}$ is still an open issue in the case of 
non-circular orbits, but it is likely that equation~(15) captures the
primary dependence on $E$ and $J$, which is sufficient for our purpose.
\footnote{The best observational evidence that equation~(15) is at 
least approximately correct is the strong inverse correlation between 
the mean densities of globular clusters in the Milky Way and their 
present galactocentric distances, $\bar\rho\propto R^{-1.7\pm0.2}$ 
(see Fig.~7 of Innanen et al. 1983). For many position-velocity 
distributions, this implies a similar, if not identical, relation 
between $\bar\rho$ and $R_p$.}
Inserting $\bar{\rho}(E,J)$ into equation~(7) gives the rate of 
evaporation by two-body relaxation $\mu_{ev}$ as a function of 
$E$ and $J$.

To compute the fractional rate of evaporation by gravitational shocks 
$\nu_{sh}$ as a function of $E$ and $J$, we average the corresponding 
rate over $R$ and $\theta$, weighting by the number of clusters with 
each of these coordinates. 
The number of clusters at any radius is proportional to the time spent 
there and hence inversely proportional to the radial component of the 
velocity there. 
The orientations of the orbits are assumed to be random.
Thus, we have
\beq
\bar{\nu}_{sh}(E,J) = \frac{2}{P_R}\int_{R_p}^{R_a} \frac{dR}{V_R(R)} 
    \int_{0}^{\pi/2} \nu_{sh}(E, J, R,\theta) \sin\theta d\theta.
\eeq
Here, the functional dependence of $\nu_{sh}$ on $E$, $J$, $R$, and 
$\theta$ follows from equation~(8) and the adiabatic correction factor 
$\bar{A}$ (see Appendix~A and Figure~12). 
The radial and azimuthal periods of the orbits are given by 
\beq
P_R = 2\int_{R_p}^{R_a} \frac{dR}{V_R(R)},
\eeq
\beq
\frac{1}{P_{\phi}} = \frac{J}{\pi P_R} \int_{R_p}^{R_a}
    \frac{dR}{R^2 V_R(R)}.
\eeq
Furthermore, the radial and vertical components of the velocity 
(relative to the disk) are given by
\beq
V_R = [2E-(J/R)^2 - 2\Phi(R)]^{1/2},
\eeq
\beq
V_Z = V_T \sin \theta = (J/R) \sin \theta,
\eeq
where $V_T$ is the transverse component of the velocity (orthogonal to
the radius in the orbital plane). Equation (20) is valid because, just
as a cluster passes through the disk, the radial and vertical components
of its velocity are orthogonal.
This simplifies our calculations substantially.
Inserting equations (8) and (20) into equation (16), we have
\beq
\bar{\nu}_{sh}(E,J) = \frac{0.0226\kappa_s} 
    {G\bar{\rho} P_R P_{\phi} J^2}
    \int_{R_p}^{R_a} \frac{R^2 g_m^2(R)}{V_R(R)} dR
    \int_{0}^{\pi/2} \frac{\bar{A}(E,J,R, \theta)}{\sin \theta}d\theta.
\eeq
When $\mu_{ev}(E,J)$ and $\bar{\nu}_{sh}(E,J)$ are substituted 
into equations (9), (10) and (2), we obtain the mass function 
$\psi(M,t; E,J)$ of clusters on orbits specified by $E$ and $J$. 

For some purposes, we are interested in how the mass function depends
on position ${\bf R}$ and velocity ${\bf V}$ rather than energy
$E$ and angular momentum $J$.
With this in mind, we define $f(M,{\bf R,V},t)dM d{\bf R}d{\bf V}$ as 
the number of clusters in the small element of mass-position-velocity
space $dM d{\bf R}d{\bf V}$ centered on $(M,{\bf R,V})$ at time $t$.
We assume for simplicity that the cluster system is spherical and
non-rotating. 
\footnote{In fact, the cluster system would develop some asphericity, 
even if it initially had none, as a consequence of the different rates 
at which clusters with different orbital orientations are disrupted 
by disk shocks.
The globular cluster systems in many galaxies consist of both a 
nearly spherical, slowly rotating (halo) component and a moderately 
flattened, rapidly rotating (disk) component.
In the Milky Way, about 27\% of the known globular clusters are 
members of the disk component [this is the fraction of clusters
in the Harris (1996, 1999) compilation with metallicities above the
disk-halo division ${\rm [Fe/H]} = -0.8$ specified by Zinn (1985)].
A complete analysis of the disruption of clusters would take these 
complications into account.
Our simple model should provide a good approximation to the mass 
function and its dependence on galactocentric radius, the main goals 
of this study, because we average the rate of evaporation by disk 
shocks over orbital orientations [see equations (16)--(21)] and 
because, for most clusters, disk shocks are not the dominant 
disruptive process.}
Thus, by the Jeans theorem, we have 
\beq
f(M,{\bf R,V},t) = \psi(M,t;E,J) f_0(E,J),
\eeq
where $f_0(E,J)$ is the initial distribution function, defined such
that $f_0(E,J)d{\bf R}d{\bf V}$ is the fraction of clusters in the 
small element of position-velocity space $d{\bf R}d{\bf V}$ with 
energies and angular momenta near $E$ and $J$ at $t=0$.
Equation~(22) implies that the disruption of clusters does not 
alter their orbits.
The mass function at the radius $R=|{\bf R}|$ is given by the 
integral of $f(M,{\bf R,V},t)$ over all velocities; using 
equation~(22) and evaluating the Jacobean for the transformation
between $(V_R, V_T)$ and $(E, J)$, we obtain
\bey
\psi(M,R,t)&=&4\pi \int_0^{\infty} dV_R \int_0^{\infty} 
              V_T f(M,{\bf R,V},t) dV_T \nonumber \\ 
           &=&\frac{4\pi}{R^2} \int_{\Phi(R)}^{\infty}dE \int_0^{J_m(E,R)} 
              \frac{J\psi(M,t;E,J)f_0(E,J)}{[2E-(J/R)^2-2\Phi(R)]^{1/2}}dJ,
\eey
where $J_m(E,R) = R [2E - 2\Phi(R)]^{1/2}$ is the maximum angular 
momentum of an orbit with a given energy $E$ and radius $R$.
Finally, we note that the average mass function in a volume bounded 
by radii $R_1$ and $R_2$ is 
\beq 
\bar{\psi}(M,t) = \frac{3}{R_2^3-R_1^3}\int_{R_1}^{R_2}\psi(M,R,t) R^2 dR. 
\eeq

In the following, we consider two simple models for the initial 
distribution function of the clusters. 
The first is the Eddington model:
\beq 
f_0(E,J) \propto \exp(-E/\sigma^2) \exp[-\onehalf (J/R_A \sigma)^2].
\eeq
This has velocity dispersions $\sigma_R=\sigma$ and 
$\sigma_T=\sigma[1+(R/R_A)^2]^{-1/2}$ in the radial and transverse 
(i.e., orthogonal) directions, respectively, where the anisotropy 
radius $R_A$ marks the transition from a nearly isotropic to a 
predominantly radial velocity distribution.
In a logarithmic potential, the initial density profile (number of 
clusters per unit volume) is 
\beq 
n_0(R) \propto [1+ (R/R_A)^2]^{-1}R^{-\gamma} 
\eeq
with $\gamma = (V_c/\sigma)^2$. 
Distribution functions like the Eddington arise frequently in 
simulations of gravitational collapse, in which violent relaxation 
is nearly complete in the inner regions but not in the outer
regions.
The second model we consider has an initial distribution function
of the form 
\beq
f_0(E,J) \propto \exp(-E/\sigma^2)  J^{-2\beta}.
\eeq
In this case, the radial and transverse velocity dispersions are
$\sigma_R=\sigma$ and $\sigma_T=\sigma(1-\beta)^{1/2}$, and in a 
logarithmic potential, the initial density profile is a power law,
\beq
n_0(R) \propto R^{-2\beta - \gamma},
\eeq
again with $\gamma = (V_c/\sigma)^2$.
We refer to this as the scale-free model. 
It is not clear which physical processes would produce a scale-free
distribution function, although gravitational clustering in a
self-similar hierarchy might do so.
For our purposes, the most important difference between the Eddington
and scale-free models is that, in the former, the velocity anisotropy 
increases outward, whereas in the latter, it is the same at all radii. 
Thus, the distribution of pericenters is narrower in the Eddington model 
than it is in the scale-free model (see the formulae in Appendix~B and 
Figure~13).

Before presenting the results of our calculations, we pause here to 
enumerate the parameters in our models. These are: 
    the escape probability parameter $\xi_e$, 
    the circular velocity of the galaxy $V_c$, 
    the central surface density and scale radius of the disk, 
        $\Sigma_d(0)$ and $R_d$, 
    the index characterizing the initial density profile $\gamma$,
    and the anisotropy radius $R_A$ (for the Eddington model) or the 
    anisotropy parameter $\beta$ (for the scale-free model).
As standard values of these parameters, we adopt 
$\xi_e=0.045$, $V_c=220$~\kms, $\Sigma_d(0)=720~\Mo~{\rm pc}^{-2}$, 
$R_d=3$~kpc, and $\gamma=2.5$, $R_A=5$~kpc (Eddington) 
or $\gamma=3.5$, $\beta = 0.5$ (scale-free).
The first of these is the escape probability parameter in the
H\'enon (1961) model, which should approximate the effective 
value of $\xi_e$ for the pre- and post-core collapse evolution 
of clusters with a realistic spectrum of stellar masses.

Our standard values of $V_c$, $\Sigma_d(0)$, and $R_d$ are appropriate 
for the disk of the Milky Way (see Binney \& Merrifield 1998).
For example, the standard values of $\Sigma_d(0)$ and $R_d$ imply 
that the surface density of the disk is $50~\Mo~{\rm pc}^{-2}$ at 
the solar position, $R=8$~kpc. 
Our standard values of $\gamma$ were chosen so that the final density
profile of the cluster system in the models would approximate the 
observed profile (see below).
We have chosen the standard value of $R_A$ to be the same as the 
median galactocentric radius of the globular clusters in the Milky 
Way, $R_h\approx 5$~kpc. 
This ensures that the velocity anisotropy at $R_h$ in the Eddington 
model is the same as that at all radii in the scale-free model, namely
$\sigma_R=\sqrt{2}\sigma_T$.
This is more radial anisotropy than appears to exist in the present
velocity distribution of globular clusters in the Milky Way (Frenk 
\& White 1980; Dinescu, Girard, \& van Altena 1999).
However, it is similar to the radial anisotropy of halo stars in the
solar neighborhood (Binney \& Merrifield 1998) and may be appropriate 
for the initial anisotropy of globular clusters (since clusters on 
elongated orbits are preferentially disrupted). 

In the following, we explore the influence of different physical 
processes on the mass function by varying the parameters with respect 
to their standard values and comparing results from the Eddington and
scale-free models.
We calculate the mass function at times up to $t=12$~Gyr,
the age of globular clusters in the Milky Way favored in several recent
studies (Gratton et al. 1997; Reid 1997; Chaboyer et al. 1998).

\section{Results}

The aim of this section is to present the results of our calculations
and to compare them with observations.
Figure~2 shows the mass functions of young star clusters in the merging
Antennae galaxies and old globular clusters in the Milky Way.
The former is based on deep {\it UBVI} images taken with the WFPC2 on 
{\it HST} and mass-to-light ratios derived from stellar population 
synthesis models (Zhang \& Fall 1999). 
The latter is based on the luminosities of the 146 clusters with 
known distances in the most recent compilation of data by Harris 
(1996, 1999) and $M/L_V=3$.
Both samples are believed to be reasonably complete over the ranges
plotted [$\log(M/\Mo) > 3.8$ and $\log(M/\Mo) > 2.9$, respectively]. 
If there is any incompleteness, however, it is likely that more 
clusters are missing from the low-mass ends of the functions.
Here, and throughout this section, we have plotted the function 
$\Psi(\log M)$, the number of clusters per unit $\log M$, against 
$\log (M/\Mo)$. 
This is related to the function $\psi(M)$ defined earlier, the number 
of clusters per unit $M$, by $\Psi(\log M) = (\log e)^{-1} M\psi(M)$.
The use of $\Psi(\log M)$ facilitates some comparisons with 
observations.
The dashed parabola in Figure~2 depicts the usual lognormal 
mass function with a peak at $M_p=2\times10^5~\Mo$ and a dispersion 
of $\sigma(\log M)=0.5$, corresponding to a Gaussian distribution of 
magnitudes with $\langle M_V \rangle=-7.3$ and $\sigma(M_V)=1.2$ 
(for $M/L_V=3$).
 
As we have already mentioned, and as Figure~2 also demonstrates,
the shapes of the mass functions of the young clusters in the 
Antennae and the old clusters in the Milky Way are very different.
The former declines monotonically as $\Psi(\log M)\propto M^{-1}$, 
whereas the latter rises to a peak at $M_p\approx2\times10^5~\Mo$ 
and then declines.
Another fact evident from Figure~2 is that the lognormal function 
provides a good representation of the empirical mass function 
(histogram) of globular clusters at high masses but not at low masses.
For $M \lesssim M_p$, the observations can be fitted better by 
$\Psi(\log M) \propto M$, corresponding to $\psi(M) = {\rm const}$. 
This behavior in the empirical mass function was first shown by
McLaughlin (1994). 
As we have pointed out here---for the first time we believe---the
form $\psi(M) = {\rm const}$ is a robust signature of evaporation
by two-body relaxation.
This behavior in the mass function can be traced to the fact that, in
the late stages of evolution, the masses of tidally limited clusters 
decrease linearly with time. 
Thus, the time $dt$ they spend in each small interval of mass $dM = 
\dot M dt$ is independent of $M$, leading to $\psi(M) = {\rm const}$.
\footnote{This is consistent with the relation $dN/dt_{rh}={\rm const}$ 
at small $t_{rh}$ noted by Hut \& Djorgovski (1992).}

The results of our calculations are presented in Figures~3--11.
In these diagrams, we plot the mass function at times $t = 0, 1.5, 
3, 6$, and 12~Gyr. 
The peak of $\Psi(\log M)$ at $t=12$~Gyr is indicated by an 
upward-pointing arrow.
We first explore the effects of different initial mass functions with 
the parameters fixed at their standard values. 
Figure~3 shows the evolution of the mass functions, averaged over all 
radii (actually, $1<R<100$~kpc) for the Eddington initial distribution
function; Figure~4 shows the corresponding results for the scale-free 
initial distribution function.
The four initial mass functions we consider are: 
  (1) a power law, $\psi_0(M)\propto M^{\beta}$ with $\beta=-2$, 
  (2) the same power law but truncated below $M=3\times 10^5~\Mo$,
  (3) a Schechter function, $\psi_0(M) \propto 
      M^{\beta}\exp(-M/M_*)$ with $\beta=-2$ and $M_*=5\times 10^6~\Mo$, 
  (4) a lognormal function, $\psi_0(M)\propto \exp\{-\onehalf
      [\log (M/M_p)/\sigma(\log M)]^2\}$ 
      with $M_p=2\times 10^5~\Mo$ and $\sigma(\log M)= 0.5$.

In all our models, the mass function develops a peak, and at 
$t=12$~Gyr, this is remarkably close to the observed peak, 
despite the very different initial conditions. 
For low-mass clusters ($M\lesssim M_p$), the disruption is dominated 
by two-body relaxation, which, as noted above, leads to $\psi(M) = 
{\rm const}$ and $\Psi(\log M)\propto M$, in excellent agreement 
with the empirical mass function (histograms). 
This is true whether the initial mass function lies above or below
the relation $\psi(M)={\rm const}$, as illustrated in the left- 
and right-hand panels of Figures 3 and 4.
Even if the formation of low-mass clusters were suppressed entirely,
they would appear later as the remnants of intermediate-mass clusters 
on their way to destruction.
High-mass clusters ($M\gtrsim M_p$) are mainly affected by stellar 
evolution and gravitational shocks.
For the reasons given above, these shift $\Psi(\log M)$ to smaller 
$\log M$ but leave its shape nearly invariant (when dynamical 
friction is neglected).
Thus, the empirical mass function can always be matched above the peak
by a suitable choice of the initial mass function, as illustrated in 
the lower panels of Figures 3 and 4, with the Schechter and lognormal 
functions.
Unfortunately, neither theory nor observation provides much guidance 
on the form of $\psi_0(M)$ at high $M$.
Because of small-number statistics, the mass function of old globular
clusters in the Milky Way is uncertain above $10^6~\Mo$, and 
that of young star clusters in the Antennae above $3\times 10^5~\Mo$ 
(see Figure~2).

We have also computed the total number of clusters $N_T(t)$ and 
their total mass $M_T(t)$ by integrating $\psi(M,t)$ and $M\psi(M,t)$ 
over $M$. 
Table~1 lists the present disruption rates, $|\dot{N}_T/N_T|$ and 
$|\dot{M}_T/M_T|$, and the present survival fractions, $N_T/N_{T0}$ 
and $M_T/M_{T0}$, for the four different initial conditions shown 
in Figure~3. 
In the two cases with steep initial mass functions, the survival 
fractions depend on the lower cutoff $M_l$ of the integration over $M$; 
thus, we list results for a range of values, $M_l=1-10^4~\Mo$.
The other entries in Table~1 are not sensitive to $M_l$.
We note that the timescales for the disruption of existing clusters,
$|\dot{N}_T/N_T|^{-1}$ and $|\dot{M}_T/M_T|^{-1}$, are comparable to 
the present age (actually $1-4$ times longer).
The reason for this is that most of the clusters with shorter disruption 
timescales have already perished.
The present values of $|\dot{N}_T/N_T|$ in our models agree with the 
total disruption rate of globular clusters in the Milky Way estimated 
by Hut \& Djorgovski (1992) from the empirical distribution of 
half-mass relaxation times ($\dot{N}_T = -5\pm3~{\rm Gyr}^{-1}$ 
and $N_T = 98$). 
Furthermore, the present values of $|\dot{M}_T/M_T|$ are similar to 
the typical (median) disruption rates of individual globular clusters
estimated by Gnedin \& Ostriker (1997).
We find that the present total number and mass of clusters, $N_T$ 
and $M_T$, are small fractions of their initial values, $N_{T0}$ 
and $M_{T0}$, especially when the initial mass function rises toward 
low masses.
The values of $M_T/M_{T0}$, in particular, indicate that a substantial
fraction of the field stars in the galactic spheroid could be the 
debris of disrupted clusters. 
However, even in the most extreme case considered here (the Schechter 
initial mass function with a lower cutoff at $M_l=1~\Mo$), the survival
fraction is a few times larger than the ratio of the mass in globular 
clusters  to the mass in the galactic spheroid (about 1\%).
Thus, to account for all the field stars in the spheroid by disrupted
clusters, the initial mass function would have to rise more steeply 
than $\psi_0(M)\propto M^{-2}$ for $M\lesssim 10^4~\Mo$. 

In Figure 5, we plot the number density profiles of the cluster 
system at different times for the Eddington and scale-free initial 
distribution functions and the Schechter initial mass function.
The profiles become flatter because clusters are destroyed faster
near the galactic center, although a nearly steady form is reached 
by $t \approx 1.5$~Gyr. 
The final profiles in both models are in reasonable, although not
perfect, agreement with the observed profile, which we have derived 
from the same compilation of data as we used for the mass function
(Harris 1996, 1999).
In fact, we chose the standard values of the parameter $\gamma =
(V_c/\sigma)^2$, after some adjustment, to achieve this match. 

Figures~6 and~7 show the evolution of the mass function for the same 
models when averaged over the inner and outer parts of the galaxy
($R<5$~kpc and $R>5$~kpc), the boundary between these being close 
to the median radius of globular clusters in the Milky Way.
In both cases, the peak mass is higher in the inner region. 
This is caused by the higher rate of two-body relaxation, resulting 
from the higher mean density, and the higher rate of gravitational 
shocks near the galactic center, the former effect being more 
important than the latter (see below).
For the Eddington model, the logarithmic difference in the peak 
mass between inner and outer clusters is $\Delta \log M_p = 0.2$, 
corresponding to a difference in mean absolute magnitudes of 
$\Delta \langle M_V \rangle = 0.5$ (for constant $M/L_V$).
For the scale-free model, the differences are $\Delta \log M_p = 
0.65$ and $\Delta \langle M_V \rangle = 1.6$.
For comparison, we find $\Delta \langle M_V \rangle = 0.16\pm0.26$ 
for the globular clusters with $R<5$~kpc and $R>5$~kpc in the Harris 
(1996, 1999) compilation. 
Thus, the radial variation of the peak mass in the Eddington model 
is consistent with observations (at the 1.3~$\sigma$ level), whereas 
that in the scale-free model is not.
Furthermore, the width of the mass function in the Eddington model 
agrees better with the observed one.
The explanation for these differences in the mass functions can be 
found in the different radial variations of the velocity anisotropy 
in the two models.
In the Eddington model, the anisotropy increases outward, causing a 
relatively narrow distribution of pericenters and hence disruption 
rates; in the scale-free model, the anisotropy is constant, causing 
a relatively wide distribution of pericenters and disruption rates
(see Appendix B).
From now on, we consider only models with the Eddington initial 
distribution function. 

The effect of changing the escape probability parameter $\xi_e$ is 
shown in Figure~8. 
As expected, the evolution of the mass function is slower and the 
peak mass is smaller for smaller $\xi_e$ and conversely for larger 
$\xi_e$.
The effective value of $\xi_e$ is not known precisely, although we
have argued above that it should be close to the H\'enon and our
standard value (0.045) when allowance is made for both the pre- and 
post-core collapse evolution of clusters with a realistic spectrum 
of stellar masses.
In fact, the similarity between the peak mass in our models and the 
observations indicates that the actual value of $\xi_e$ cannot differ 
from our standard value by more than a factor of about two.

In Figure~9, we show the effects of varying the velocity anisotropy
radius $R_A$ on the mass function of clusters in the inner and outer 
parts of the galaxy ($R<5$~kpc and $R>5$~kpc). 
Small values of $R_A$ imply predominantly radial orbits, with mostly 
small pericenters and hence large mean densities, whereas large values 
of $R_A$ imply a nearly isotropic velocity distribution, with a wide 
range of pericenters and mean densities. 
This is why the mass function evolves faster and the peak mass is 
larger for smaller $R_A$.
Moreover, the more radial are the orbits, the more similar are the 
peak masses in the inner and outer parts of the galaxy.
For $R_A=2.5$~kpc, we find $\Delta \log M_p\approx 0$, whereas for 
$R_A=15$~kpc, we find $\Delta \log M_p= 0.5$.
These correspond, respectively, to $\Delta \langle M_V\rangle \approx 0$ 
and $1.2$ (for constant $M/L_V$).
The former is consistent with the observed value, $\Delta \langle M_V 
\rangle = 0.16\pm0.26$, whereas the latter is probably not. 
Thus, a substantial degree of radial anisotropy is required in the 
initial velocity distribution for consistency with the weak radial 
variation in the empirical mass function. 
The present velocity distribution may in fact be nearly isotropic
(Frenk \& White 1980), but as a result of the preferential disruption
of clusters on elongated orbits, the initial distribution would have 
been more anisotropic.

Figure~10 shows the effects of altering the surface density of the 
disk. 
Here, we present results for an exponential disk with double the 
standard central density, i.e., $\Sigma_d(0)=1440~\Mo {\rm pc}^{-2}$, 
and for no disk at all. 
Figure~10 also indicates how our results depend on the energy-mass 
conversion factor $\kappa_s$ since the rate of evaporation by 
gravitational shocks is proportional to $\kappa_s \Sigma_d^2$.
As expected, a more massive disk causes the mass function of star 
clusters to evolve faster.
However, in this case, unlike two-body relaxation, the peak mass 
decreases.
The reason for this is that, as noted above, gravitational shocks 
shift the mass function to smaller masses while leaving its shape
nearly invariant.
However, gravitational shocks are less important than two-body
relaxation in the disruption of low-mass clusters, even allowing 
for possible uncertainties in $\kappa_s$ and $\Sigma_d$.
This conclusion is also supported by the N-body models of Vesperini 
\& Heggie (1997).
Thus, the peak mass and its radial variation and the low-mass 
shape of the mass function are all determined primarily by two-body
relaxation rather than gravitational shocks, contrary to 
some statements in the literature.

All our previous results were computed for a galaxy like the 
Milky Way, with $V_c = 220$~\kms.
It is also of interest to know how the mass function would evolve 
in other galaxies, with different circular velocities.
For this purpose, we assume that the masses and sizes of galaxies
scale as $M_g\propto V_c^k$ and $R_g \propto V_c^{k-2}$ (to satisfy
the virial theorem, $V_c^2 \propto M_g/R_g$).
Then the mean internal densities of star clusters, like those of
their host galaxies, scale as $\bar\rho\propto M_g/R_g^3 \propto 
V_c^{6-2k}$.
Recent estimates of the exponent in the baryonic Tully-Fisher
relation lie in the range $k\approx$~3--4 (Bell \& de Jong 2001,
and references therein).
For $k$ at the lower end of the range, the mass function of star
clusters is independent of $V_c$ (since $\bar\rho$ is independent
of $V_c$).
For $k$ at the upper end of the range, the mass function has the
dependence on $V_c$ shown in Figure~11.
In this case, the peak mass decreases by $\Delta \log M_p = 0.5$ 
($\Delta \langle M_V \rangle = 1.2$) as $V_c$ increases from 110
to $440$~\kms.
This is probably larger than allowed by observations (Harris 1991).
More definitive comparisons will require better knowledge of 
the relations between $M_g$, $R_g$, and $V_c$ and possibly more
complicated models for the internal structure of galaxies (e.g.,
with finite core radii).
Moreover, dynamical friction, which is neglected in our models,
may be important in galaxies with small $V_c$.

\section{Comparison With Previous Studies}

Several other researchers have suggested that disruptive processes 
would cause the mass function of star clusters to evolve toward 
something like a lognormal function. 
Okazaki \& Tosa (1995) based their analysis on the survival 
triangle in the mass-radius (i.e., $\log M$-$\log r_h$) plane, 
defined by setting the characteristic timescales for evaporation 
by two-body relaxation and gravitational shocks, $t_{ev}$ and 
$t_{sh}$, equal to the current time $t$ (Fall \& Rees 1977).
Okazaki \& Tosa assumed that clusters inside the triangle would 
exist without any loss of mass ($M=M_0$ for $t_{ev}>t$ and 
$t_{sh}>t$), while clusters outside the triangle would not exist 
at all ($M=0$ for $t_{ev}<t$ or $t_{sh}<t$).
In other words, the mass of each cluster was assumed to be a step 
function of time, with the step at $t=\min(t_{ev}, t_{sh})$, rather 
than to decrease continuously with time as shown in Figure~1.
The clusters were postulated to have a power-law initial mass function
and a Gaussian initial distribution
of $\mu \equiv \log (M/r_h^{\beta})$, with $\beta=2.6$ or $4.1$.
The mass function at later times was then obtained from this by 
integrating over $r_h$ with the limits of integration set by the 
survival triangle.
In this approach, the shape of the present mass function 
is determined largely by the assumed shape of the initial 
distribution of $\mu$, which is not explained.
Ostriker \& Gnedin (1997) followed the same approach in a study of 
the radial variation of the mass function, except that they postulated 
a bivariate Gaussian initial distribution of $x\equiv \log (M/r_h^3)$ 
and $y\equiv \log M$ and integrated over the survival triangle in 
these coordinates, including the side for dynamical friction.

The evolution of the mass function of star clusters by disruptive 
processes was also considered by Elmegreen \& Efremov (1997). 
They claimed this evolution would take the form $\psi(M,t) = 
\psi_0(M) \exp [-\nu(M)t]$ with $\nu(M)=\dot{M}/M$.
Unfortunately, this is not correct, as one may verify by direct 
substitution into equation~(1). 
Elmegreen \& Efremov also claimed the disruption rate would have
a strong inverse dependence on mass, $\nu\propto M^{-\gamma}$
with $\gamma \approx 2$. 
This is based on the current disruption rates of surviving clusters 
estimated by Gnedin \& Ostriker (1997).
However, since the correlation between $\nu$ and $M$ is relatively 
weak, the value of $\gamma$ estimated in this way is quite uncertain 
(see Figure~2 of Elmegreen \ Efremov 1997).
In fact, the data appear to be equally consistent with $\nu\propto
M^{-1}$, the relation expected for the disruption of tidally limited
clusters by two-body relaxation [see equations~(6) and~(7) here].
The mass function proposed by Elmegreen \& Efremov has a peak at 
a ${\rm few}\times 10^5~\Mo$ and approaches the initial mass function 
for higher masses, but its shape for lower masses, where disruption 
is important, differs markedly from the solutions presented here.

Murali \& Weinberg (1997a,b,c) studied the disruption of star 
clusters by two-body relaxation and gravitational shocks using 
a series of Fokker-Planck models. 
They followed the evolution of cluster systems with a halo component
alone and with both halo and disk components.
The initial distribution functions of the halo and disk components
were represented by the scale-free model (called the Mestel sphere)
and the Mestel disk, respectively, while the initial mass function 
was represented by a truncated power law. 
Murali \& Weinberg (1997c) found that this model, with suitable
choices of parameters, could reproduce many of the observed 
properties of the globular cluster system in the Milky Way.
However, with the Fokker-Planck approach, they could only follow 
the evolution of clusters on a relatively sparse grid ($5\times4
\times5$) in the variables $R_a$, $M$, and $J/J_c$.
The four mass bins covered the range $1\times 10^5 < M < 5\times 
10^6~\Mo$, thereby excluding the low-mass clusters most susceptible 
to disruption.
In any case, since Murali \& Weinberg did not display the mass 
function at later times, we cannot make useful comparisons between
our results and theirs.

Vesperini (1997, 1998) used analytical and semi-analytical models 
to study the evolution of the mass function of star clusters 
resulting from two-body relaxation, gravitational shocks, stellar 
evolution, and dynamical friction. 
He assumed the clusters were tidally limited and on circular orbits 
perpendicular to the galactic disk. 
The initial mass function was assumed to be a truncated power law or 
a lognormal function.
Vesperini found in many cases that the final mass function in his
models resembled the empirical mass function of old globular clusters. 
However, for the truncated power law, the peak of the final mass 
function was well below the observed peak unless the truncation mass 
was large, i.e., $ M_l \gtrsim 10^5~\Mo$.
In addition, the mass functions in Vesperini's models have a strong
dependence on galactocentric radius because, with all the clusters 
on circular orbits, no radial mixing occurs.
As we have shown here, the low-mass end of the empirical mass function
of globular clusters can be reproduced even if the initial mass 
function has no truncation (i.e., with $M_l =0$). 
Moreover, we find that radial mixing of orbits is necessary to account
for the weak radial dependence of the empirical mass function.
Vesperini speculated that the lognormal mass function represented a 
quasi-equilibrium distribution.
We find instead that the high-mass shape of the mass function, 
whatever its initial form, is approximately preserved, while the
low-mass shape, $\psi(M) = {\rm const}$, is flatter than the
lognormal function.
Vesperini (2000, 2001) has recently used his models to predict
the radial variation and dependence of the mass function of star
clusters on the properties of their host galaxies (although the
clusters are still assumed to be on circular orbits).

Baumgardt (1998) studied the evolution of the mass function of 
star clusters resulting from two-body relaxation and dynamical 
friction but not gravitational shocks or stellar evolution.
He assumed the clusters were tidally limited at the pericenters 
of their orbits and approximated their disruption by means of a
simple analytical model. 
Baumgardt computed the orbits of the clusters numerically, with 
energies and angular momenta drawn from an initial distribution 
function similar, but not identical, to our scale-free model 
(note that $\gamma$ in his notation is $\gamma + 2\beta$ in 
our notation).
He adopted a power-law initial mass function with $\beta = -2$ 
and a broad initial distribution of half-mass radii. 
Baumgardt found that the mass function developed a peak and that, 
after a Hubble time, this coincided roughly with the peak in the 
empirical mass function of old globular clusters. 
He also found that the peak mass was larger at smaller radii, 
with $\Delta \log M_p \approx 0.4$ ($\Delta \langle M_V \rangle
\approx 1.0$) for clusters inside and outside $R=10$~kpc, and he 
noted that this was probably incompatible with observations. 
Our results for the scale-free model generally agree with 
Baumgardt's, although detailed comparisons are difficult because 
his mass functions are very noisy. 
As we have shown here, the  mass function in the Eddington model
has  a weaker radial variation than that in the scale-free model,
in satisfactory agreement with observations.

\section{Discussion}

We have presented a series of simple, largely analytical models to 
compute the effects of disruption on the mass function of star clusters. 
Our models include evaporation by two-body relaxation and 
gravitational shocks and mass loss by stellar evolution.
A virtue of our approach is that it leads to a clear understanding of 
how each disruptive process shapes the mass function of star clusters.
Our goal has been to determine under what initial conditions the mass 
function evolves into a form resembling that of old globular clusters.
We make two idealizations to simplify our calculations.
First, we neglect correlations between the effects of two-body 
relaxation and gravitational shocks.
A comparison with more accurate Monte Carlo, Fokker-Planck, and N-body 
models indicates that the errors introduced by this approximation are 
acceptably small, especially for low- and high-mass clusters.
Second, we assume the galactic potentials in which the clusters move
are static and spherical.
This ensures that the pericenter of each orbit remains fixed.
We discuss below how our results would be modified if the galactic
potentials were time-dependent and/or non-spherical.

We find that, for a wide variety of initial conditions, the mass 
function in our models develops a turnover or peak, which, after 
12 Gyr, is remarkably close to the observed peak in the mass function 
of globular clusters ($M_p\approx2\times10^5~\Mo$).
Below the peak, the evolution is dominated by two-body relaxation, 
and the mass function always develops a tail of the form $\psi(M) =
{\rm const}$. 
This reflects the linear decrease in the masses of tidally limited
clusters with time just before they are destroyed.
The predicted form of $\psi(M)$ at and below the peak coincides well
with the observed form.
We interpret this as strong support for the idea that evaporation by 
two-body relaxation played a major role in shaping the mass function 
of globular clusters at low and intermediate masses ($M \lesssim M_p$).
Above the peak, the evolution of the mass function is dominated by 
stellar evolution at early times and by gravitational shocks at late 
times (when dynamical friction is neglected).
These processes operate at fractional rates that are independent of 
the masses of the clusters and hence tend to preserve the shape of 
the mass function at high masses (in a log-log plot).
We also find that the disruption of clusters can contribute 
substantially to the field star population in the galactic 
spheroid if the initial mass function of the clusters rises 
steeply toward low masses. 

The radial variation of the mass function within a galaxy depends 
on the initial position-velocity distribution of the clusters.
The reason for this is that the rate of evaporation by two-body 
relaxation and gravitational shocks depends on the orbits of the 
clusters, especially their pericenters.
If most of the orbits are circular or if the velocity distribution 
is isotropic, the mass function will vary more with galactocentric
radius than is observed.
This variation can be reduced by the greater radial mixing that occurs 
when the velocity distribution has some radial anisotropy.
However, to obtain a nearly uniform mass function within a galaxy, 
the radial anisotropy must increase outward, producing a distribution 
of pericenters that is narrower than the distribution of instantaneous 
positions of the clusters.
We illustrate this by our models with Eddington and scale-free initial
distribution functions, both of which have the same anisotropy at the
median radius of the globular cluster system.
In the former, the radial variation of the mass function is compatible
with observations, whereas in the latter, it is not.
Unfortunately, the initial position-velocity distribution of globular 
clusters is not known because most of the original clusters have been 
destroyed.
However, since the destruction occurs preferentially for clusters on 
elongated orbits, the initial velocity distribution must have been
more anisotropic than the present one.

Our conclusions to this point are based on models with static, 
spherical galactic potentials. 
In such models, each cluster returns to the same pericenter on each 
revolution about the galaxy.
In galaxies with non-spherical potentials, however, the pericenter of
a cluster may change from one revolution to the next. 
This effect should help to dilute the radial variation of the mass 
function.
An even more effective mixing of pericenters and consequent 
homogenization of the mass function may occur in galaxies with
time-dependent potentials. 
Large variations in the potential are a natural consequence of the
formation and evolution of galaxies by hierarchical clustering.
Each time one galaxy merges with another, the orbits of the clusters 
are likely to be scrambled to some degree by violent relaxation. 
In this way, the mass functions of clusters in the inner and outer 
parts of the galaxies would also be combined or averaged, erasing
any prior radial variations.
In the hierarchical picture, merging is expected to be important 
early in the histories of all galaxies and late in the histories of 
some galaxies.
It is not clear, however, whether merging occurred recently enough 
in most galaxies to account for the observed uniformity of the mass 
functions of their globular clusters.
It is doubtful, for example, that the Milky Way or Andromeda galaxies
experienced any major mergers in the last 8 Gyr or so (otherwise, 
their old disks would have been disrupted).
Nevertheless, the mass functions of globular clusters in galaxies with
time-dependent and/or non-spherical potentials should have less radial 
variation than those in the idealized models presented here.

Our models and some of those mentioned in the previous section have 
several observational implications.
The first is that the peak of the mass function of clusters should
increase with age. 
This might be observable in galaxies in which clusters formed 
continuously over long periods of time. 
Alternatively, the evolution of the peak mass might be observable 
in galaxies with bursts of cluster formation at different times,
such as in a sequence of merger remnants. 
This test may be difficult, however, because the luminosity 
corresponding to the peak mass is relatively small for young clusters 
(since $M_p$ varies more rapidly with $t$ than $M/L_V$ does).
The second observational implication is that the peak mass should 
decrease with increasing distance from the centers of galaxies, unless 
this has been completely diluted by the mixing of pericenters
discussed above. 
Searches for radial variations in the peak mass have so far been 
inconclusive. 
This test is difficult because the diffuse light of the galaxies also 
varies with radius, making it harder to find faint clusters in the inner 
regions. 
Finally, the strong dependence of the peak mass on the ages of clusters
and the weak dependence on their positions within and among galaxies 
cast some doubt on the use of the peak luminosity as a standard candle 
for distance estimates. 
This method may be viable, however, if the samples of clusters are
carefully chosen from similar locations in similar galaxies.

Our models also have implications for attempts to understand the 
formation of star clusters of different types. 
The shape of the mass function above the peak is largely preserved 
as clusters are disrupted and hence should reflect processes at the 
time they formed.
Below the peak, however, the shape of the mass function is determined
entirely by disruption, mainly driven by two-body relaxation, and hence 
contains no information about how the clusters formed. 
If there were any feature in the initial mass function, such as a 
Jeans-type lower cutoff, it would have been erased.
In our models, the only feature in the present mass function, the peak 
at $2\times 10^5~\Mo$, is largely determined by the condition that 
clusters of this mass have a timescale for disruption comparable to 
the Hubble time. 
Thus, it is conceivable that star clusters of different types (open, 
populous, globular, etc) formed by the same physical processes with 
the same initial mass function and that the differences in their 
present mass functions reflect only their different ages and local 
environments, primarily the strength of the galactic tidal field. 
Our results therefore support the suggestion that at least some of 
the star clusters formed in merging galaxies can be regarded as young 
globular clusters. 
Further investigations of these objects may shed light on the 
processes by which old globular clusters formed.

\acknowledgements

We thank Howard Bond, Kenneth Freeman, Douglas Heggie, Massimo 
Stiavelli, Bradley Whitmore, Chun Xu, and an anonymous referee for 
helpful discussions and correspondence.
This work was supported in part by the National Aeronautics and Space
Administration through grant number GO-07468 from the Space Telescope
Science Institute and by the National Science Foundation through grant
number PHY94-07194 to the Institute for Theoretical Physics.

\appendix

\section{Adiabatic Correction Factors}

In this appendix, we present approximate expressions for the average
adiabatic corrections to the energy changes within a cluster 
(relative to those for an impulsive response). 
This is important because the stars in the inner region of a cluster 
may respond nearly adiabatically to a gravitational shock, while those
in the outer region may respond nearly impulsively. 
The local adiabatic correction factors for first- and second-order 
energy changes, $A_1(x)$ and $A_2(x)$, are usually regarded as 
functions of the dimensionless variable $x=\omega(r)\tau_{sh}$, where 
$\omega(r)$ is the orbital angular frequency at a radius $r$ within
the cluster, $\tau_{sh}=H/V_Z$ is the effective duration of the shock,
and $H$ is the scale height of the disk. 
Since these energy changes are caused by tidal accelerations, they
are proportional to $r^2$. 
Thus, the mass-weighted average adiabatic correction factors for 
first- and second-order changes in the total energy of a cluster 
are given by 
\beq
\bar{A}_{1,2} = \frac{\int_0^{r_t}  r^2 A_{1,2}[x(r)]\rho(r)r^2 dr}
{\int_0^{r_t} r^2\rho(r)r^2 dr}.
\eeq

We compute $\bar{A}_{1,2}$ from equation~(A1) with the following 
approximations. 
Several formulae have been proposed for the local adiabatic correction 
factors (see Gnedin et al. 1999 for a summary). 
We adopt the limiting form derived by Weinberg (1994) from linear
perturbation theory:
\beq
A_1(x)=A_2(x) = (1+x^2)^{-3/2}.
\eeq
This approximates the results of N-body simulations for slow shocks. 
Similar expressions, but with more negative exponents, apply for fast
shocks (Gnedin \& Ostriker 1999).
The errors we introduce by using equation~(A2) in both cases are 
acceptably small since $A_1$ and $A_2$ both approach $1$ for fast shocks.

The effects of gravitational shocks are greatest at large radii 
($x\lesssim 1$), where the internal potential of a cluster,
usually very centrally concentrated, is nearly Keplerian.
Thus, we approximate the orbital angular frequency by the formula 
\beq
\omega(r)=\left(\frac{r}{r_t}\right)^{-3/2}\left(\frac{GM}{r_t^3}\right)^{1/2}.
\eeq
This is exact for circular orbits near the tidal radius. 
The coefficient in equation~(A3) would be larger for radial orbits 
with apocenters near $r_t$ ($\sqrt{2}$ rather than 1), but the stars
would then spend much of their time at smaller radii 
[rms radius $=(5/8)^{1/2}r_t$]. 
These effects largely cancel, indicating that equation~(A3) is
approximately valid for most stars in the outer region of a cluster 
(where $A_1$ and $A_2$ are non-negligible).

Finally, in the numerical integrations of equation~(A1), 
we approximate the density profile of the clusters by the simple formula
\beq
\rho(r)\propto \frac{1}{r^2}\left(1-\frac{r}{r_t}\right)^{5/2}.
\eeq
In the outer region, this matches the profile of the King (1966)
model, $\rho(r)\propto (1/r-1/r_t)^{5/2}$ for $r\rightarrow r_t$. 
Furthermore, in the inner region, it has the singular 
behavior appropriate for core collapse models, $\rho(r) \propto r^{-2}$ 
for $r\rightarrow 0$ (Spitzer 1987).
The half-mass radius of our model is $r_h=0.18r_t$, reasonably close 
to that of the H\'enon (1961) model ($r_h=0.15r_t$). 
Figure~12 shows the resulting average adiabatic correction factor, 
$\bar{A} = \bar{A_1} = \bar{A_2}$, as a function of the dimensionless 
variable 
\beq
x_t \equiv \omega(r_t)\tau_{sh}=(4\pi G\bar{\rho}/3)^{1/2}(H/V_Z).
\eeq
This depends on $E$ and $J$ through $\bar{\rho}$ [see equations~(14) 
and (15)] and on $J$, $R$, and $\theta$ through $V_Z$ [see equation~(20)].
We adopt $H=260$~pc in all our calculations (Binney \& Merrifield 1998).

\section{Distribution of Pericenters}

In this appendix, we derive expressions for the density of pericenter
distances $n(R_p)$ from the distribution function $f(E,J)$, the density 
of clusters in position-velocity space with orbital energies and angular 
momenta near $E$ and $J$.
To do so, we introduce the auxiliary function $N(E,J)$, defined such 
that $N(E,J)dEdJ$ is the number of clusters with energies and angular 
momenta in the small intervals $(E,E+dE)$ and $(J,J+dJ)$. 
Then the number of clusters with pericenters in the interval 
$(R_p,R_p+dR_p)$ is given by 
\beq
4\pi R_p^2 n(R_p) dR_p = \int_{R_pV_c}^{\infty} N[E(R_p,J),J]
   \left|\left(\frac{\partial E}{\partial R_p}\right)_J\right| dJdR_p,
\eeq
where $E(R_p,J)$ is the energy of an orbit with pericenter $R_p$ and
angular momentum $J$. 
Using equation~(14) and the relation $N(E,J)=8\pi^2 Jf(E,J)P_R(E,J)$ 
(see equation 2-89 of Spitzer 1987), where $P_R(E, J)$ is the 
radial period defined in equation~(17), we obtain
\beq
n(R_p)=\frac{2\pi}{R_p^3}\int_{R_pV_c}^{\infty} Jf[E(R_p,J),J]
       P_R[E(R_p,J),J] [(J/R_p)^2-V_c^2(R_p)] dJ.
\eeq

As in the main text, we assume the galaxy has a logarithmic potential, 
with $V_c =~{\rm const}$. 
Then, for the Eddington initial distribution function, we obtain
\beq
n_0(R_p)\propto \pi V_c^3 R_p^{-\gamma}
 \int_1^{\infty} \exp(-\onehalf\gamma x[1+(R_p/R_A)^2])I(x)(x-1) dx,
\eeq
and for the scale-free initial distribution function
\beq
n_0(R_p)\propto \pi V_c^{3-2\beta} R_p^{-2\beta-\gamma}
   \int_1^{\infty} \exp(-\onehalf\gamma x)x^{-\beta}I(x)(x-1) dx.
\eeq
Here, we have changed the variable of integration from $J$ to 
$x=(J/R_pV_c)^2$ and introduced the parameter $\gamma=(V_c/\sigma)^2$ 
(as before) and the function
\beq
I(x)=\int_{1}^{y(x)} [x(z-1)-z\ln z]^{-1/2} dz,
\eeq
where $y(x)$ is the upper root of the equation 
\beq x(y-1)=y\ln y.\eeq

In Figure 13, we plot the initial density of pericenters $n_0(R_p)$, 
along with the initial density of cluster positions $n_0(R)$, for 
both the Eddington and scale-free initial distribution functions 
with the standard values of the parameters ($\gamma=2.5$ and $R_A=5$~kpc 
for the former, $\gamma=3.5$ and $\beta=0.5$ for the latter). 
This shows the important result that, for the Eddington model, the 
distribution of pericenters is narrower than the distribution of
cluster positions, while for the scale-free model, the distributions
of pericenters and cluster positions have the same power-law form.
Consequently, there is less radial variation in the disruption rates 
for the Eddington distribution function than for the scale-free 
distribution function.

\newpage

\begin{figure}
\centerline{\resizebox{6in}{!}{\rotatebox{0}{\includegraphics{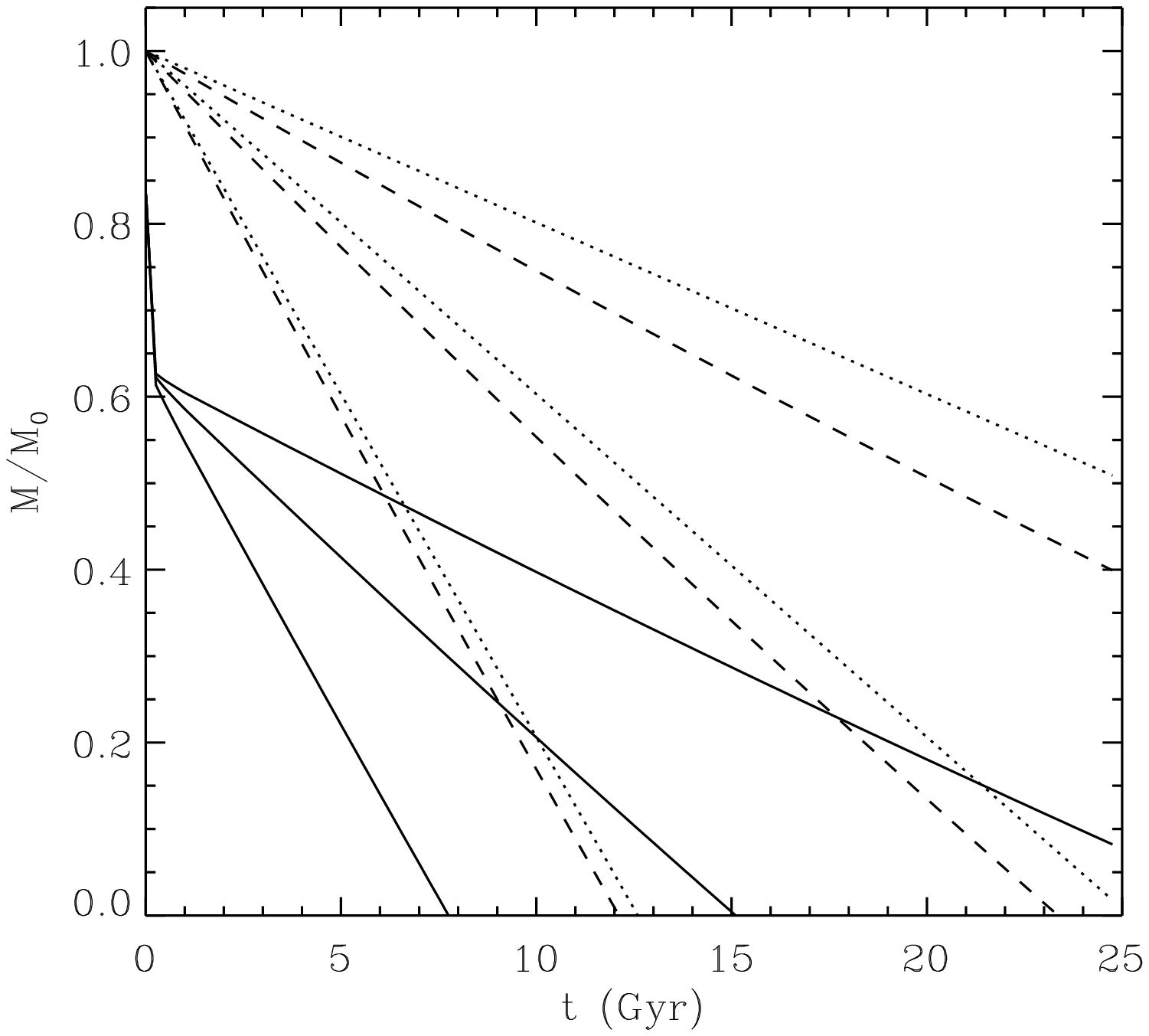}}}}
\caption{
Evolution of the masses of clusters predicted by equations~(7)--(10) 
with three different initial masses: $M_0 = (1, 2, 4) \times10^5~\Mo$.
All three clusters have the same orbit, with $R_p = 4$~kpc, $R_a = 
6$~kpc, and $\theta = 45^\circ$.
The dotted lines show the evolution with two-body relaxation
alone, the dashed lines with two-body relaxation and gravitational
shocks, and the solid lines with two-body relaxation, gravitational
shocks, and stellar evolution.
Note that stellar evolution dominates in the early stages, that
gravitational shocks become relatively less important as the mass
decreases, and that two-body relaxation dominates in the late
stages. 
}
\end{figure}

\begin{figure}
\centerline{\resizebox{6in}{!}{\rotatebox{0}{\includegraphics{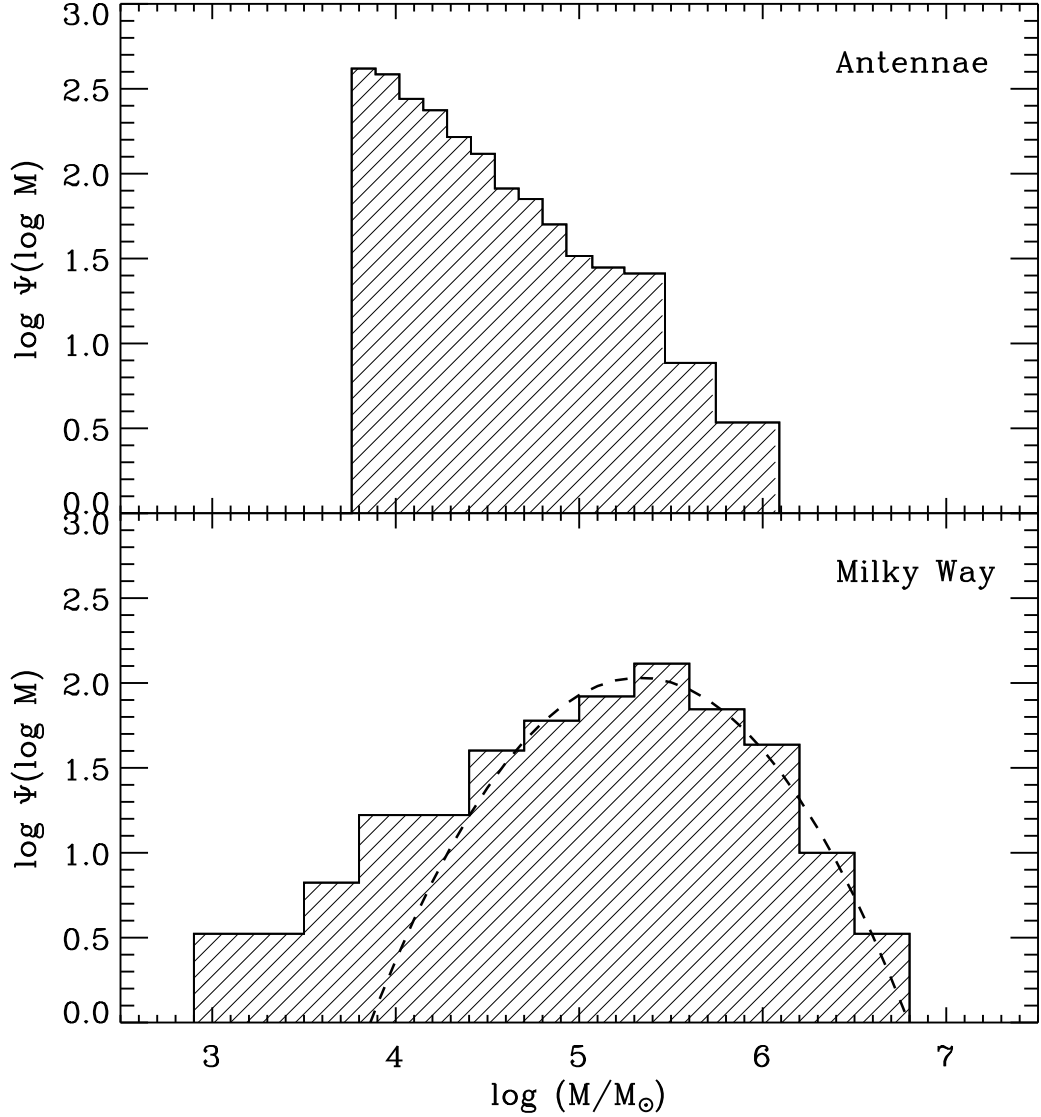}}}}
\caption{
Histograms of the masses of young star clusters in the Antennae galaxies
and old globular clusters in the Milky Way. 
The former is from Zhang \& Fall (1999); the latter is based on data 
compiled by Harris (1996, 1999). 
Note that the mass function of the young clusters declines monotonically 
over the entire observed range, $\log(M/\Mo)>3.8$, whereas the mass 
function of the old clusters rises to a peak and then declines. 
The dashed curve is the usual lognormal representation of the mass function 
with $M_p = 2\times10^5~\Mo$ and $\sigma(\log M)=0.5$, corresponding to a 
Gaussian distribution of absolute magnitudes with $\langle M_V \rangle = 
-7.3$ and $\sigma(M_V) = 1.2$ (for $M/L_V = 3$). 
Note that the empirical mass function of the old clusters (histogram) 
is shallower below the peak than the lognormal function.
}
\end{figure}

\begin{figure}
\centerline{\resizebox{6in}{!}{\rotatebox{90}{\includegraphics{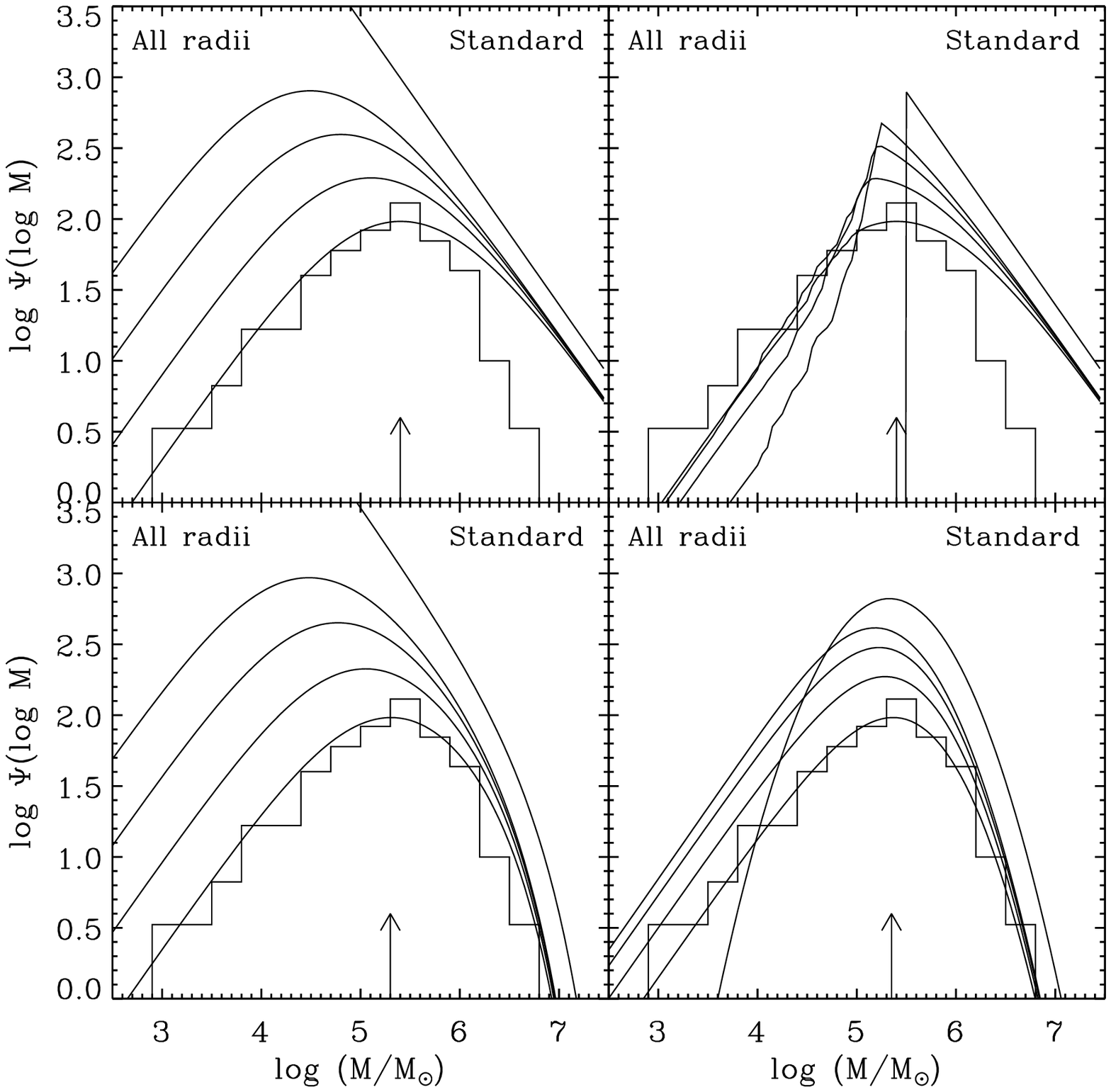}}}}
\caption{  
Evolution of the mass function, averaged over all radii, for the
Eddington initial distribution function with the 
standard parameters and different initial mass functions. 
These are: a power law of index $\beta=-2$ ({\it top left}), 
the same power law truncated at $3\times 10^5~\Mo$ ({\it top right}), 
a Schechter function with $\beta=-2$ and $M_{*}=5\times10^6~\Mo$ 
({\it bottom left}), a lognormal function with $M_p=2 \times 
10^5~\Mo$ and $\sigma(\log M)=0.5$ ({\it bottom right}).
Each mass function is plotted at $t=0$, 1.5, 3, 6, and 12~Gyr;
the arrows indicate the peak at $t=12$~Gyr. 
The histograms depict the empirical mass function of globular clusters in 
the Milky Way. 
Note that the peak mass in the models is similar to that in the 
observations for the four different initial conditions.
}
\end{figure}

\begin{figure}
\centerline{\resizebox{6in}{!}{\rotatebox{90}{\includegraphics{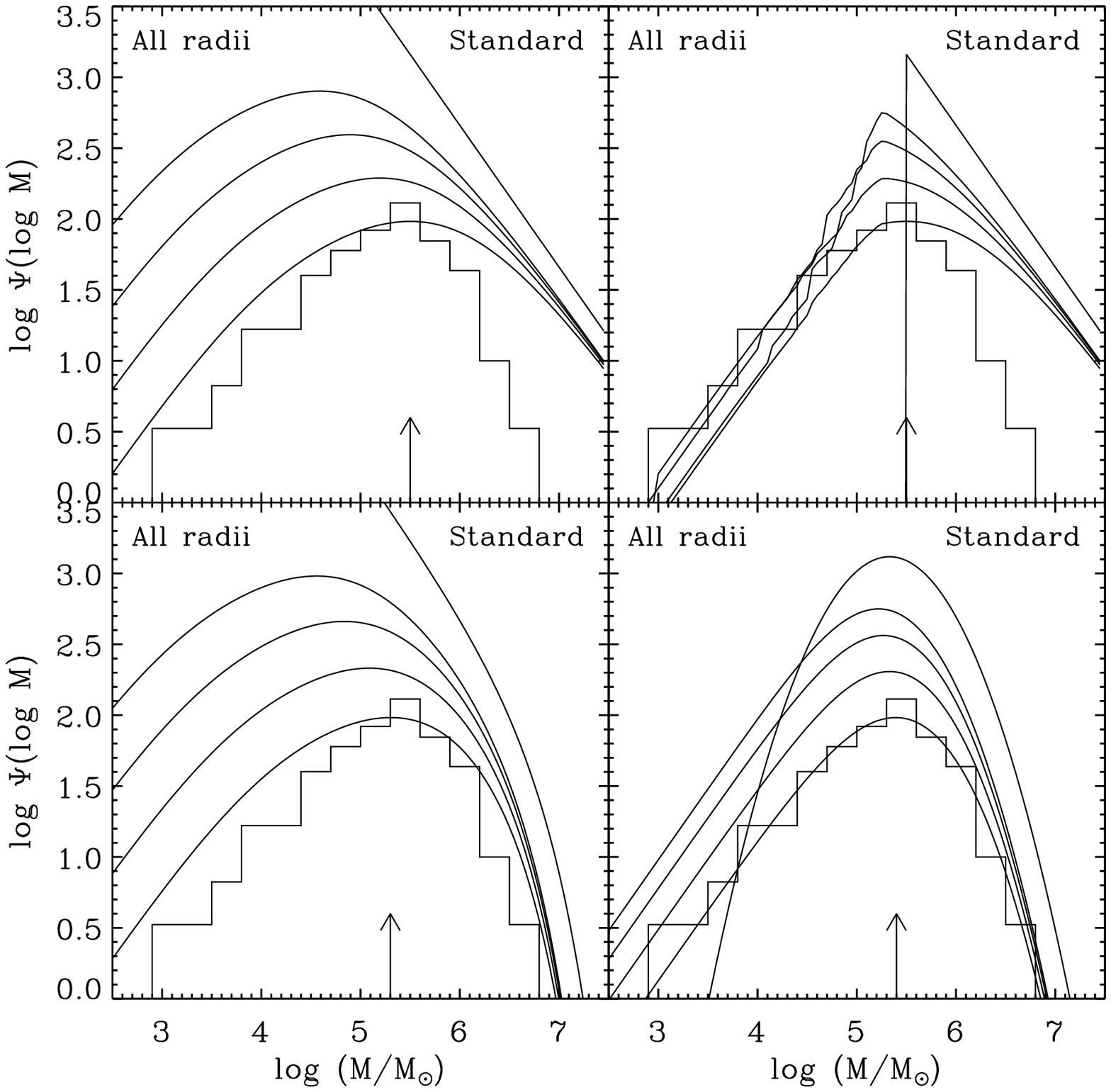}}}}
\caption{  
Evolution of the mass function, averaged over all radii, for the
scale-free initial distribution function with the standard parameters 
and different initial mass functions. 
These are: a power law of index $\beta=-2$ ({\it top left}), 
the same power law truncated at $3\times 10^5~\Mo$ ({\it top right}), 
a Schechter function with $\beta=-2$ and $M_{*}=5\times10^6~\Mo$ 
({\it bottom left}), a lognormal function with $M_p=2 \times 
10^5~\Mo$ and $\sigma(\log M)=0.5$ ({\it bottom right}).
Each mass function is plotted at $t=0$, 1.5, 3, 6, and 12~Gyr;
the arrows indicate the peak at $t=12$~Gyr. 
The histograms depict the empirical mass function of globular clusters 
in the Milky Way. 
Note that the peak mass in the models is similar to that in the
observations for the four different initial conditions. 
}
\end{figure}

\begin{figure}
\centerline{\resizebox{6in}{!}{\rotatebox{0}{\includegraphics{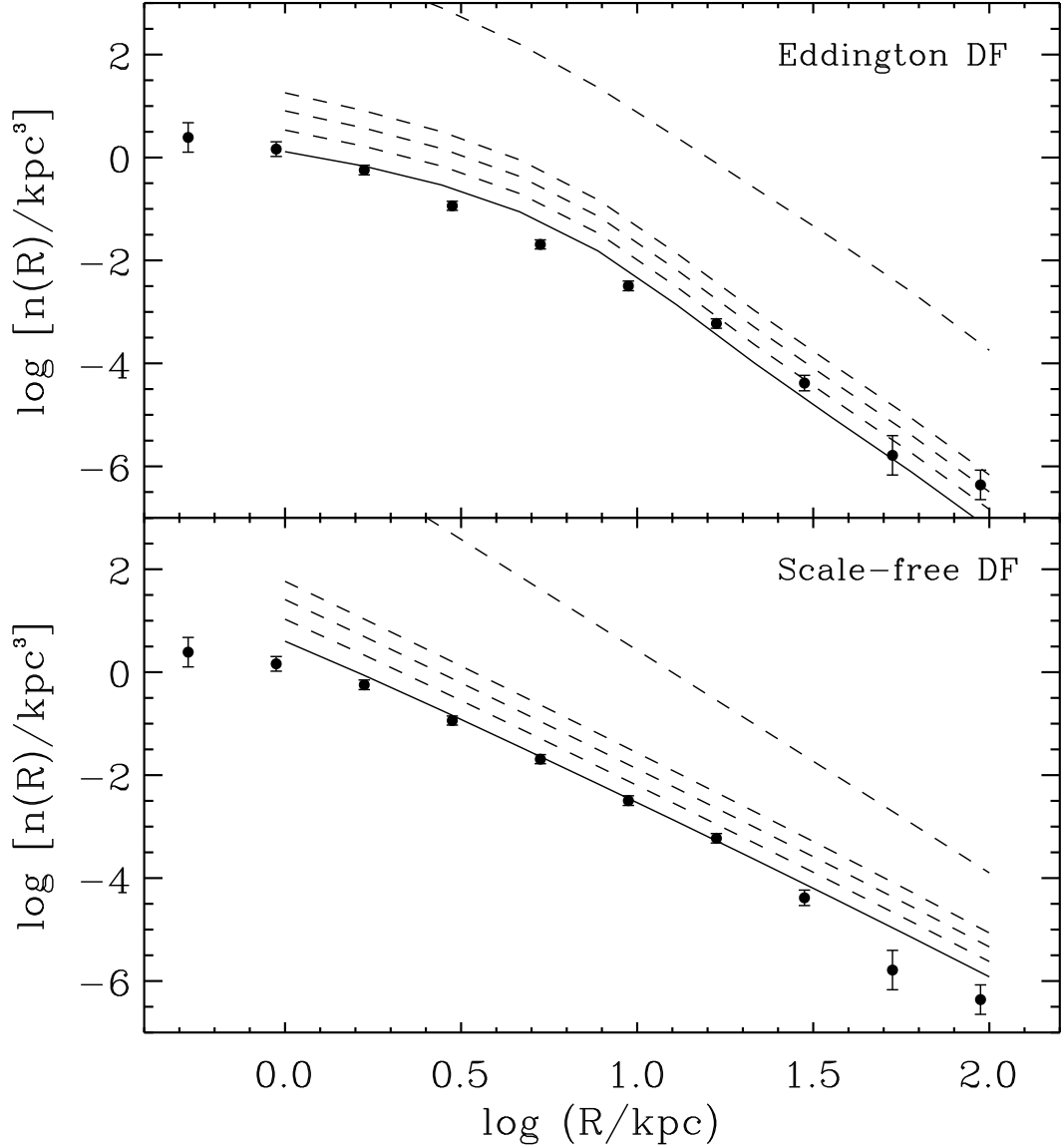}}}}
\caption{  
Evolution of the number density profile of the cluster system for
the Eddington and scale-free initial distribution functions with the 
standard parameters and the Schechter initial mass function. 
The profiles are plotted at $t=0$, 1.5, 3, 6, and 12~Gyr. 
To avoid a divergence in the initial density profile, 
the mass function is truncated at $100~\Mo$.
The data points depict the empirical profile for globular clusters
in the Milky Way.
Note that the final profiles in the models are in reasonable agreement
with the empirical profile.
}
\end{figure}

\begin{figure}
\centerline{\resizebox{6in}{!}{\rotatebox{90}{\includegraphics{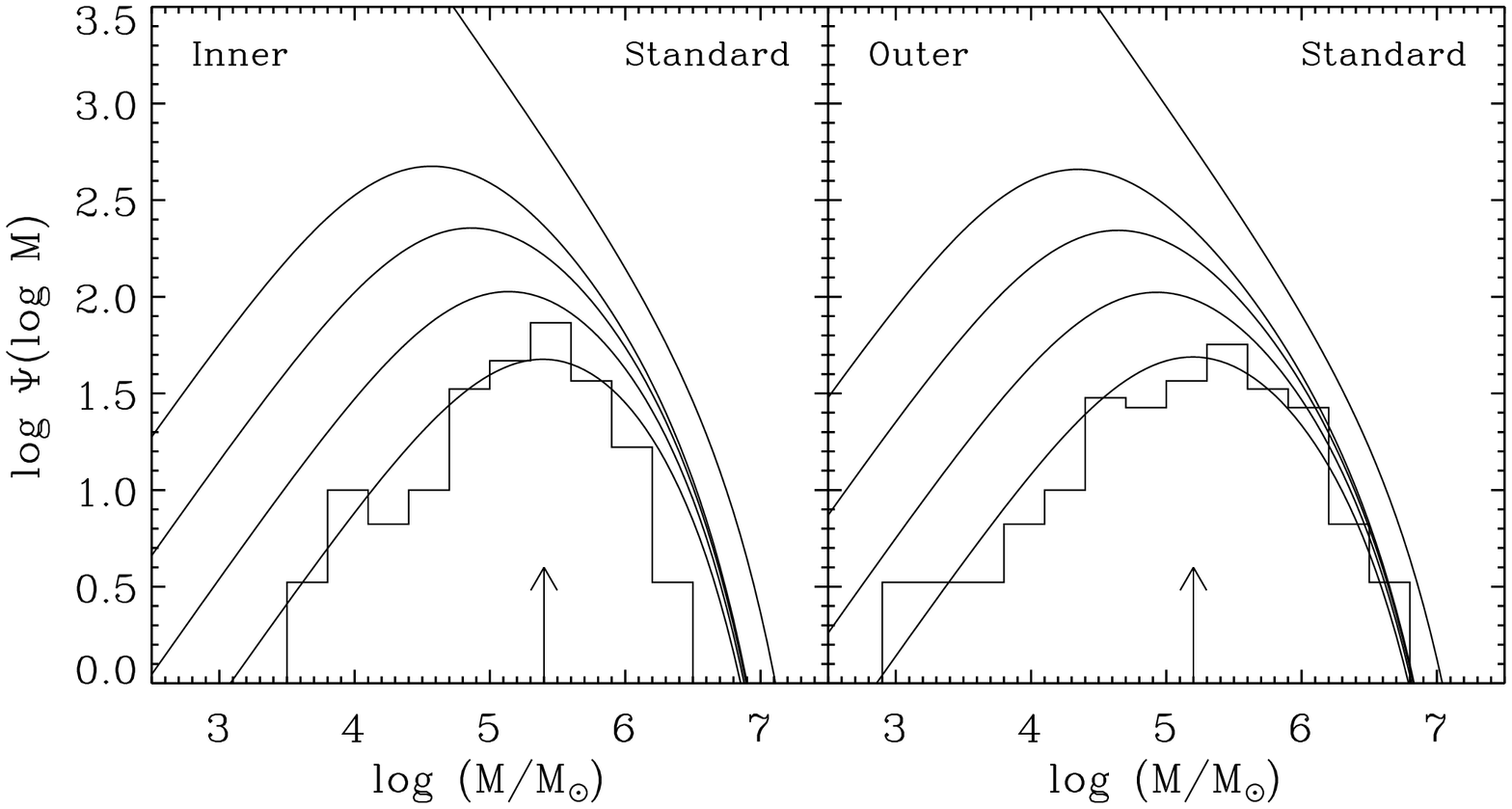}}}}
\caption{  
Evolution of the mass function, averaged over inner radii ($R < 5$~kpc)
and outer radii ($R > 5$~kpc), for the Eddington initial distribution 
function with the standard parameters and the Schechter initial mass 
function.
Each mass function is plotted at $t=0$, 1.5, 3, 6, and 12~Gyr;
the arrows indicate the peak at $t=12$~Gyr. 
The histograms depict the empirical mass functions of globular clusters 
in the Milky Way in the corresponding ranges of radii. 
Note that the shift in the peak mass in the models between inner and outer 
radii is relatively small.
}
\end{figure}

\begin{figure}
\centerline{\resizebox{6in}{!}{\rotatebox{90}{\includegraphics{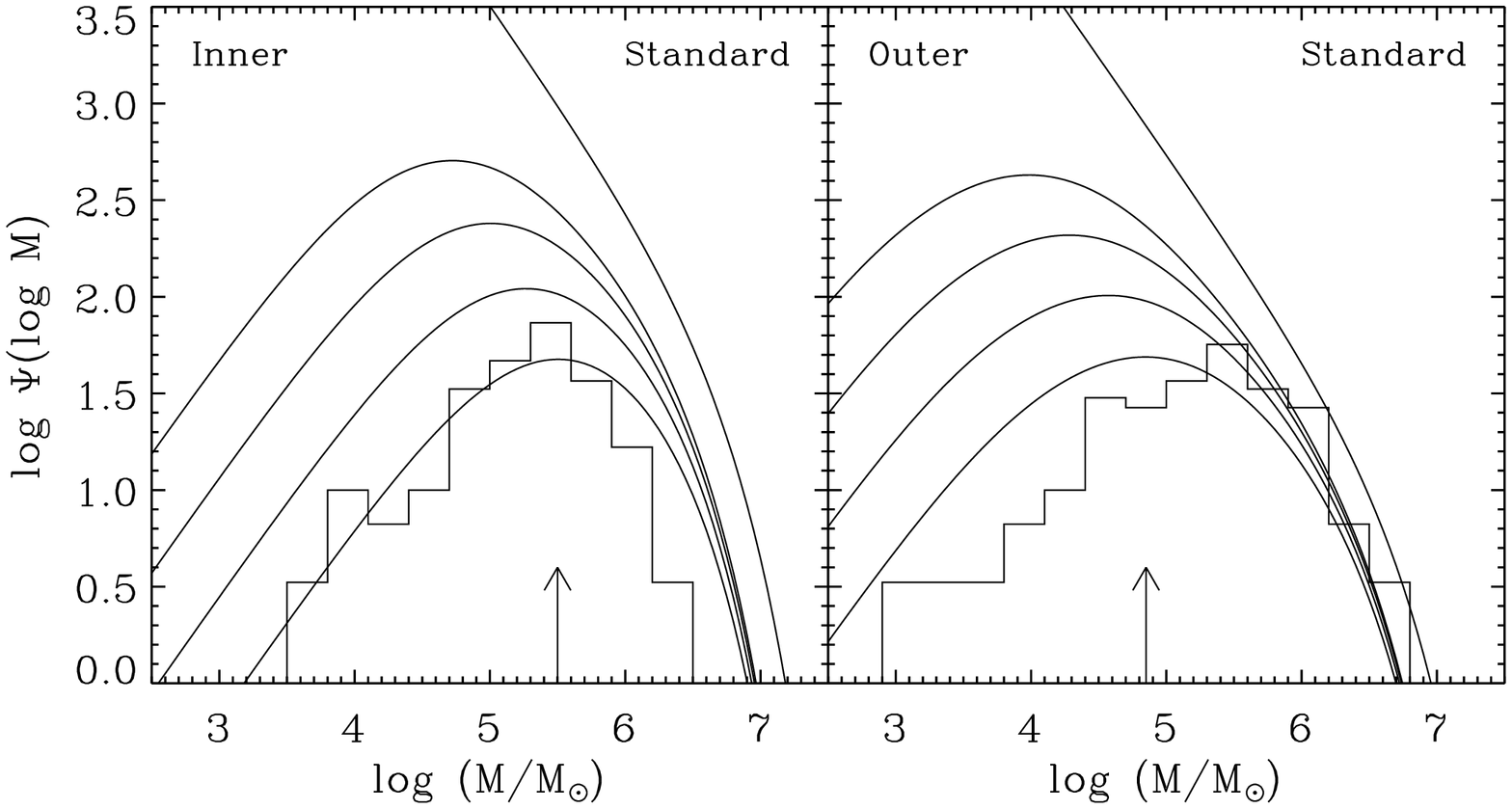}}}}
\caption{  
Evolution of the mass function, averaged over inner radii ($R < 5$~kpc)
and outer radii ($R > 5$~kpc), for the scale-free initial distribution 
function with the standard parameters and the Schechter initial mass 
function.
Each mass function is plotted at $t=0$, 1.5, 3, 6, and 12~Gyr;
the arrows indicate the peak at $t=12$~Gyr. 
The histograms depict the empirical mass functions of globular clusters 
in the Milky Way in the corresponding ranges of radii. 
Note that the shift in the peak mass in the models between inner and outer 
radii is relatively large.
}
\end{figure}

\begin{figure}
\centerline{\resizebox{6in}{!}{\rotatebox{90}{\includegraphics{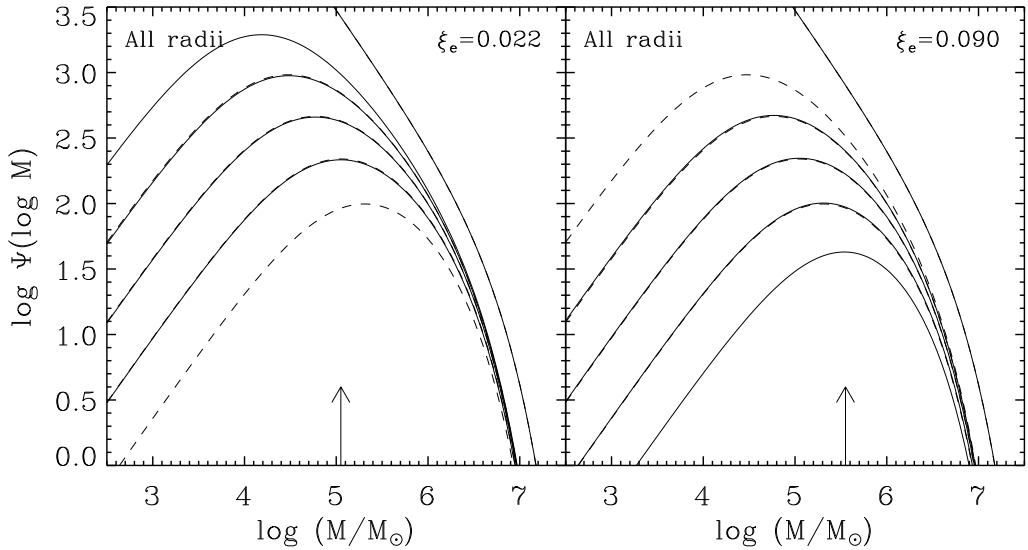}}}}
\caption{  
Evolution of the mass function, averaged over all radii, with different 
values of the escape probability parameter $\xi_e$ (as indicated), for 
the Eddington initial distribution function and the Schechter initial 
mass function.
Each mass function is plotted at $t=0$, 1.5, 3, 6, and 12~Gyr;
the arrows indicate the peak at $t=12$~Gyr. 
The dashed curves represent the same models with the standard parameters.
Note that the peak mass in the models is larger for larger $\xi_e$.
}
\end{figure}

\begin{figure}
\centerline{\resizebox{6in}{!}{\rotatebox{90}{\includegraphics{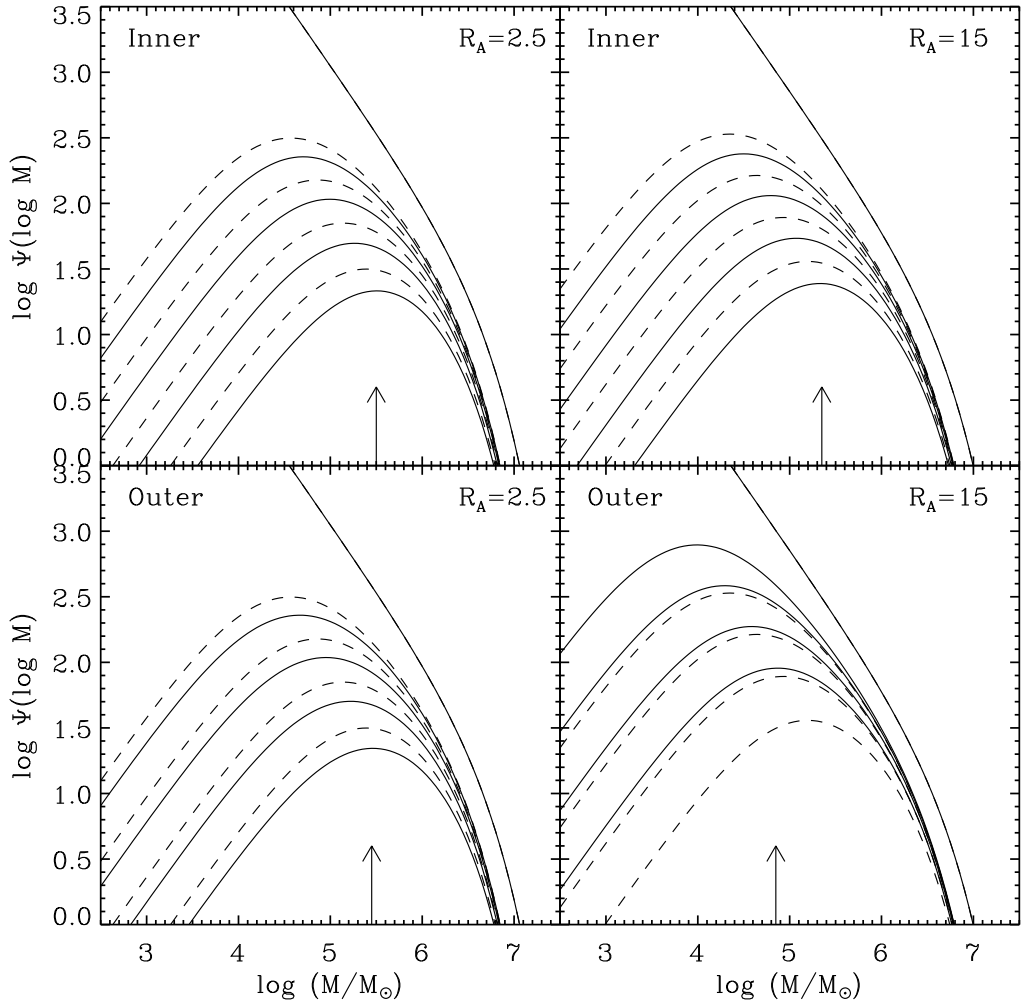}}}}
\caption{  
Evolution of the mass function, averaged over inner radii ($R < 5$~kpc) 
and outer radii ($R > 5$~kpc), with different values of the velocity 
anisotropy radius $R_A$ (in kpc, as indicated), for the Eddington initial 
distribution function and the Schechter initial mass function.
Each mass function is plotted at $t=0$, 1.5, 3, 6, and 12~Gyr;
the arrows indicate the peak at $t=12$~Gyr. 
The dashed curves represent the same models with the standard parameters.
Note that the shift in the peak mass in the models between inner and
outer radii is larger for larger $R_A$.
}
\end{figure}

\begin{figure}
\centerline{\resizebox{6in}{!}{\rotatebox{90}{\includegraphics{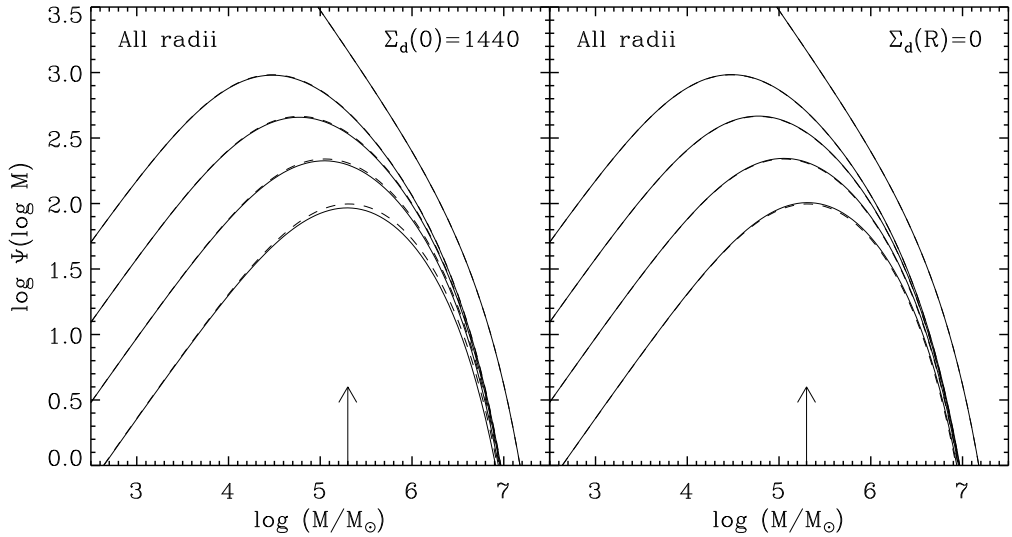}}}}
\caption{  
Evolution of the mass function, averaged over all radii, with a more 
massive exponential disk ({\it left panel}) and with no disk 
({\it right panel}), for the Eddington initial distribution 
function and the Schechter initial mass function.
Each mass function is plotted at $t=0$, 1.5, 3, 6, and 12~Gyr; 
the arrows indicate the peak at $t=12$~Gyr. 
The dashed curves represent the same models with the standard parameters.
Note that the peak mass in the models is smaller for stronger disks.
}
\end{figure}

\begin{figure}
\centerline{\resizebox{6in}{!}{\rotatebox{90}{\includegraphics{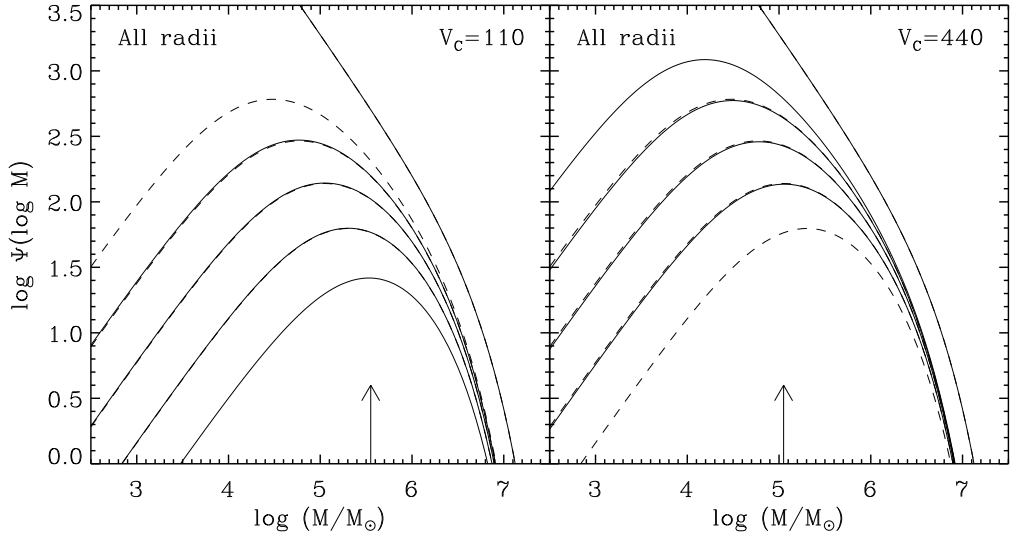}}}}
\caption{  
Evolution of the mass function, averaged over all radii, with different 
values of the circular velocity $V_c$ (in \kms,\ as indicated), for 
the Eddington initial distribution function and the Schechter initial 
mass function, and for $k=4$ in the relation $M_g \propto V_c^k$.
Each mass function is plotted at $t=0$, 1.5, 3, 6, and 12~Gyr;
the arrows indicate the peak at $t=12$~Gyr. 
The dashed curves represent the same models with the standard parameters.
For $k < 4$, the mass function has a weaker dependence on $V_c$ than
shown here (as explained in the text).
}
\end{figure}

\begin{figure}
\centerline{\resizebox{6in}{!}{\rotatebox{0}{\includegraphics{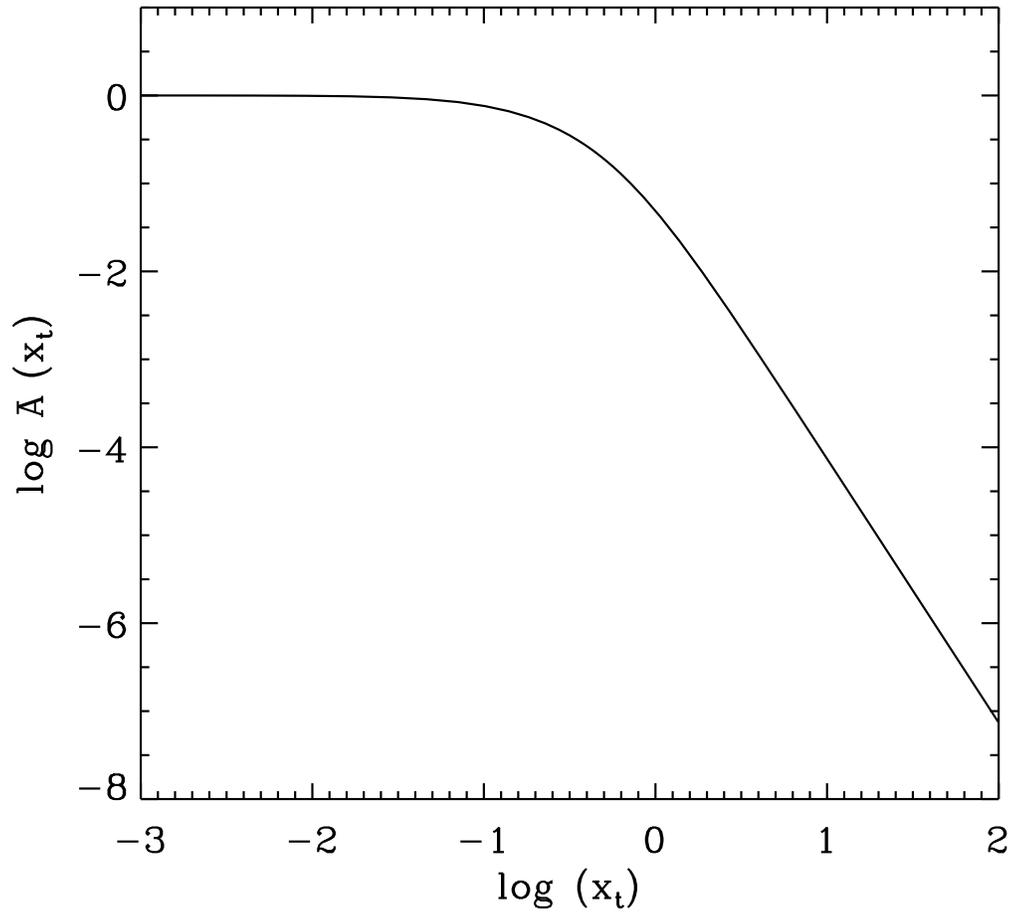}}}}
\caption{  
Average adiabatic correction factor for first- and second-order 
changes in the total energy of a cluster plotted against the 
dimensionless variable $x_t = \omega(r_t)\tau_{sh}$.
See Appendix~A for details.
}
\end{figure}

\begin{figure}
\centerline{\resizebox{6in}{!}{\rotatebox{0}{\includegraphics{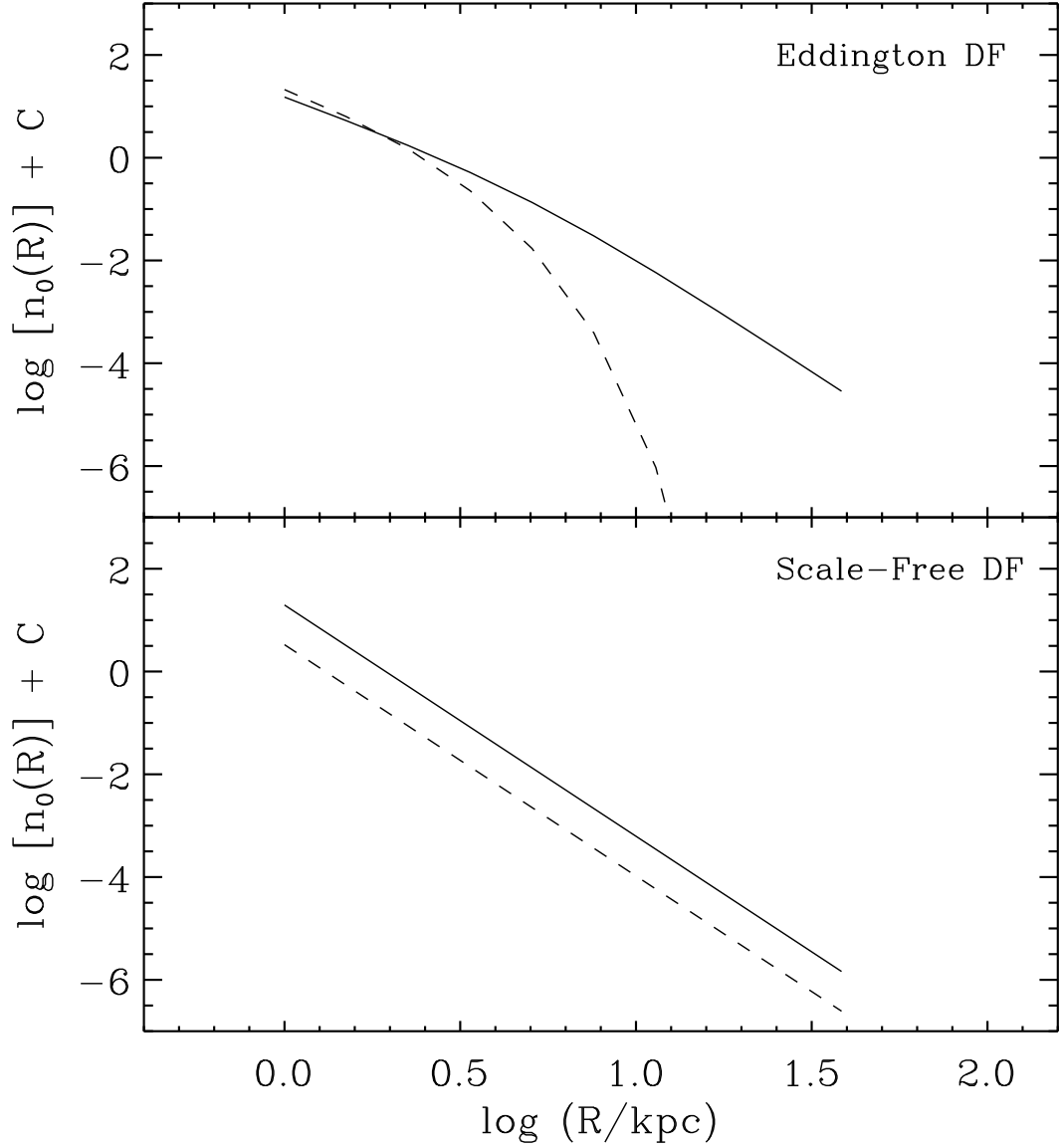}}}}
\caption{  
Initial densities of cluster positions (solid lines) and pericenters
(dashed lines) for the Eddington and scale-free distribution 
functions. 
Note that the distribution of pericenters is narrower than the
distribution of cluster positions for the Eddington model but
not for the scale-free model.
}
\end{figure}

\clearpage

\begin{deluxetable}{lllll}
\tablecaption{Present Total Disruption Rates and Survival Fractions}
\tablewidth{0pt}
\tablehead{
Initial Mass Function & $|\dot{N}_T/N_T|^{\rm a}$ & 
$|\dot{M}_T/M_T|^{\rm a}$ & $N_T/N_{T0})^{\rm b}$ & 
$M_T/M_{T0})^{\rm b}$ 
}
\startdata
Power Law           & 0.064 & 0.022 & 0.00--0.02 & 0.08--0.19 \\
Truncated Power Law & 0.051 & 0.021 & 0.47       & 0.37 \\
Schechter           & 0.079 & 0.049 & 0.00--0.01 & 0.04--0.10 \\
Lognormal           & 0.072 & 0.057 & 0.17       & 0.16  \\
\enddata

\tablenotetext{a}{Rates are in Gyr$^{-1}$.}
\tablenotetext{b}{Ranges are for a lower cutoff in $\psi_0(M)$ 
from $M_l = 1~\Mo$ to $10^4~\Mo$.}
\end{deluxetable}


\begin{thebibliography}{}

\bibitem[]{} Aguilar, L., Hut, P., \& Ostriker, J. P. 1988, \apj, 335, 720
\bibitem[]{} Baumgardt, H. 1998, \aap, 330, 480
\bibitem[]{} Bell, E. F., \& de Jong, R. S. 2001, \apj, 550, 212
\bibitem[]{} Binney, J., \& Merrifield, M. 1998, Galactic Astronomy 
             (Princeton: Princeton Univ. Press)
\bibitem[]{} Binney, J., \& Tremaine, S. 1987, Galactic Dynamics (Princeton: 
             Princeton Univ. Press)
\bibitem[]{} Caputo, F., \& Castellani, V. 1984, \mnras, 207, 185
\bibitem[]{} Chaboyer, B., Demarque, P., Kernan, P. J., \& Krauss, L. M. 1998, 
             \apj, 494, 96
\bibitem[]{} Chernoff, D. F., Kochanek, C. S., \& Shapiro, S. L. 1986, \apj, 
             309, 183
\bibitem[]{} Chernoff, D. F., \& Shapiro, S. L. 1987, \apj, 322, 113
\bibitem[]{} Chernoff, D. F., \& Weinberg, M. D. 1990, \apj, 351, 121
\bibitem[]{} Christian, C. A., \& Schommer, R. A. 1988, \aj, 95, 704
\bibitem[]{} Dickey, J. M., \& Garwood, R. W. 1989, \apj, 341, 201
\bibitem[]{} Dinescu, D. I., Girard, T. M., \& van Altena, W. F. 1999, \aj, 
             117, 1792
\bibitem[]{} Elmegreen, B. G., \& Efremov, Y. N. 1997, \apj, 480, 235
\bibitem[]{} Elson, R. A. W., \& Fall, S. M. 1985, \pasp, 97, 692
\bibitem[]{} Fall, S. M., \& Rees, M. J. 1977, \mnras, 181, 37P
\bibitem[]{} ---------. 1985, \apj, 298, 18
\bibitem[]{} Frenk, C. S., \& White, S. D. M. 1980, \mnras, 193, 295
\bibitem[]{} Fritze-von Alvensleben, U. 1999, \aap, 342, L25
\bibitem[]{} Fukushige, T., \& Heggie, D. C. 1995, \mnras, 276, 206
\bibitem[]{} Gnedin, O. Y., Lee, H. M., \& Ostriker, J. P. 1999, \apj, 522, 
             935
\bibitem[]{} Gnedin, O. Y., \& Ostriker, J. P. 1997, \apj, 474, 223
\bibitem[]{} ---------. 1999, \apj, 513, 626
\bibitem[]{} Gratton, R. G., Fusi Pecci, F., Carretta, E., Clementini, G.,
             Corsi, C. E., \& Lattanzi, M. 1997, \apj, 491, 749
\bibitem[]{} Harris, W. E. 1991, \araa, 29, 543
\bibitem[]{} ---------. 1996, \aj, 112, 1487
\bibitem[]{} ---------. 1999, http://www.physics.mcmaster.ca/Globular.html
             (revision of June 22, 1999)
\bibitem[]{} H\'enon, M. 1961, Ann. d'Astrophys., 24, 369
\bibitem[]{} Hut, P., \& Djorgovski, S. 1992, \nat, 359, 806
\bibitem[]{} Innanen, K. A., Harris, W. E., \& Webbink, R. F. 1983, \aj, 88, 
             338
\bibitem[]{} Johnstone, D. 1993, \aj, 105, 155
\bibitem[]{} Kang, H., Shapiro, P. R., Fall, S. M., \& Rees, M. J. 1990,
             \apj, 363, 488
\bibitem[]{} King, I. R. 1962, \aj, 67, 471
\bibitem[]{} ---------. 1966, \aj, 71, 64
\bibitem[]{} Kundi\'c, T., \& Ostriker, J. P. 1995, \apj, 438, 702
\bibitem[]{} Lee, H. M., \& Goodman, J. 1995, \apj, 443, 109
\bibitem[]{} Lee, H. M., \& Ostriker, J. P. 1987, \apj, 322, 123
\bibitem[]{} Leitherer, C., Schaerer, D., Goldader, J. D., Gonz\'alez Delgado,
             R. M., Robert, C., Foo Kune, D., de Mello, D. F., Devost, D., \& 
             Heckman, T. M. 1999, \apjs, 123, 3
\bibitem[]{} McLaughlin, D. E. 1994, PASP, 106, 47
\bibitem[]{} Meurer, G. R. 1995, \nat, 375, 742
\bibitem[]{} Murali, C., \& Weinberg, M. D. 1997a, \mnras, 288, 749
\bibitem[]{} ---------. 1997b, \mnras, 288, 767
\bibitem[]{} ---------. 1997c, \mnras, 291, 717
\bibitem[]{} Okazaki, T., \& Tosa, M. 1995, \mnras, 274, 48
\bibitem[]{} Ostriker, J. P., \& Gnedin, O. Y. 1997, \apj, 487, 667
\bibitem[]{} Ostriker, J. P., Spitzer, L., \& Chevalier, R. A. 1972, \apj,
             176, L51
\bibitem[]{} Reid, I. N. 1997, \aj, 114, 161
\bibitem[]{} Solomon, P. M., \& Rivolo, A. R. 1989, \apj, 339, 919
\bibitem[]{} Spitzer, L. 1987, Dynamical Evolution of Globular Clusters 
             (Princeton: Princeton Univ. Press)
\bibitem[]{} Spitzer, L., \& Chevalier, R. A. 1973, \apj, 183, 565
\bibitem[]{} van den Bergh, S., \& Lafontaine, A. 1984, \aj, 89, 1822
\bibitem[]{} Vesperini, E. 1997, \mnras, 287, 915
\bibitem[]{} ---------. 1998, \mnras, 299, 1019
\bibitem[]{} ---------. 2000, \mnras, 318, 841
\bibitem[]{} ---------. 2001, \mnras, 322, 247
\bibitem[]{} Vesperini, E., \& Heggie, D. C. 1997, \mnras, 289, 898
\bibitem[]{} Weinberg, M. D. 1994, \aj, 108, 1403
\bibitem[]{} Whitmore, B. C. 1997, in The Extragalactic Distance Scale, 
             ed. M. Livio, M. Donahue, \& N. Panagia (Cambridge: Cambridge
             Univ. Press), 254
\bibitem[]{} Whitmore, B. C., Zhang, Q., Leitherer, C., Fall, S. M., 
             Schweizer, F., \& Miller, B. W. 1999, \aj, 118, 1551
\bibitem[]{} Zhang, Q., \& Fall, S. M. 1999, \apj, 527, L81
\bibitem[]{} Zinn, R. 1985, \apj, 293, 424

\end{thebibliography}
\end{document}